\documentclass[]{article}

\usepackage{amsmath}
\usepackage{amssymb}
\usepackage{tikz}
\usepackage{graphicx}
\usepackage[footnotesize,hang]{caption}
\usepackage{pifont}
\usepackage{subfig}
\usepackage{color}
\usepackage{rotating}
\usepackage{mathtools}
\usepackage{xcolor}
\usepackage{epstopdf}
\usepackage{cite}
\usepackage{url}
\usetikzlibrary{decorations.pathreplacing}

\usepackage[margin=0.5in]{geometry}
\def\Integer{\mathbb{Z}}

\usetikzlibrary{decorations.pathmorphing}

\renewcommand{\i}{\mathrm{i}}
\newcommand{\e}{\mathrm{e}}

\renewcommand{\d}{\,\mathrm{d}}
\newcommand{\diff}[2]{\frac{\mathrm{d} #1}{\mathrm{d} #2}}

\def\XXint#1#2#3{{\setbox0=\hbox{$#1{#2#3}{\int}$}
\vcenter{\hbox{$#2#3$}}\kern-.5\wd0}}

\title{Nanoptera in nonlinear woodpile chains with zero precompression}
\author{G. Deng$^1$\footnote{Corresponding Author. Electronic address: guo.deng@mq.edu.au}, C. J. Lustri$^1$\footnote{Electronic address: christopher.lustri@mq.edu.au}}
\date{%
    $^1$Department of Mathematics and Statistics, 12 Wally's Walk, Macquarie University, New South Wales 2109, Australia\\[2ex]%
}                                

\begin{document}
\maketitle
\abstract{
We use exponential asymptotics to study travelling waves in woodpile systems modelled as singularly perturbed granular chains with zero precompression and small mass ratio. These systems are strongly nonlinear, and there is no analytic expression for their leading-order solution. We instead obtain an approximated leading-order solution using a hybrid numerical-analytic method. We show that travelling waves in these nonlinear woodpile systems are typically ``nanoptera'', or travelling waves with exponentially small but non-decaying oscillatory tails which appear as a Stokes curve is crossed. We demonstrate that travelling wave solutions in the zero precompression regime contain two Stokes curves, and hence two sets of tailing oscillations in the solution. We calculate the behaviour of these oscillations explicitly, and show that there exist system configurations which cause the oscillations to cancel entirely, producing solitary wave behaviour. We then study the behaviour of travelling waves in woodpile chains as precompression is increased, and show that there exists a value of the precompression above which the two Stokes curves coalesce into a single curve, meaning that cancellation of the tailing oscillations no longer occurs. This is consistent with previous studies, which showed that cancellation does not occur in chains with strong precompression.
}

%%%%%

\section{Introduction}\label{intro}

\subsection{Motivation}
A granular chain is composed of tightly-packed aligned particles which are solid, spherical, and frictionless, and which deform elastically upon contact. In practice, the repelling force from the contact area of adjacent spherical particles is determined by the compression between two particles. In this paper, we apply exponential asymptotics to study woodpile chains, or chains of orthogonally stacked rigid cylinders, which can be modelled as singularly perturbed granular chains.

Previous studies~\cite{Deng2021,Kim,Xu} on woodpile chains showed that typical travelling-wave solutions in  woodpile chains are ``nanoptera", which are superpositions of an exponentially localized solitary wave and non-decaying oscillations on one or both sides of the localized wave. The work of~\cite{Xu} showed that for woodpile chains without precompression, which are strongly nonlinear, there exists sets of system parameters for which the oscillations vanish by considering the Fourier transform of a woodpile chain . These parameter configurations are known as ``anti-resonance conditions". In contrast,~\cite{Deng2021} applied exponential asymptotics to show that for woodpile chains with large precompression, which are weakly nonlinear, the oscillations are never absent and there exists no anti-resonance conditions.

The purpose of this study is to  calculate an explicit asymptotic form of the oscillations in woodpile chains with zero precompression, and to compute anti-resonance conditions. We will then explain why anti-resonance conditions are not possible in systems with strong precompression by studying how the system varies as the precompression is increased, and we will demonstrate that there exists a critical value of the precompression above which anti-resonance conditions cannot occur for any choice of system parameters. 

\subsection{Granular chains}
The motion of particles in a granular chain is governed by the following system of differential--difference equations:
\begin{equation}
	m(n)\ddot{x}(n,t)=\phi'(x(n+1,t)-x(n,t))-\phi'(x(n,t)-x(n-1,t))\,,
\label{e:lattice}
\end{equation}
where $n\in\Integer$, the quantity $m(n)$ is the mass of the $n$-th particle, $x(n,t)$ is the position of the $n$-th particle at time $t$, a dot denotes differentiation with respect to time, a prime denotes differentiation with respect to space, and the interaction potential between adjacent particles is
\begin{eqnarray}
	\phi(r)=
\begin{cases}
	c(\Delta-r)^{\alpha + 1}\,, &r\leq\Delta \cr 0\,, &r>0 \end{cases}\,,~\quad c = \mathrm{constant}\,,
\label{e:potential}
\end{eqnarray}
where $\alpha > 1$ and $\Delta$ is the equilibrium overlap of adjacent particles due to the precompression generated by an external force. The choice $\alpha = 3/2$ produces a granular system with Hertzian interaction potential, known as a Hertzian chain. The interaction potential \eqref{e:potential} is zero when adjacent particles lose contact with each other, and it cannot take negative values. For algebraic convenience, we set $c=1/(\alpha+1)$ in all subsequent examples.

The interaction potential in~\eqref{e:potential} is superquadratic, as the the exponent $\alpha + 1$ is larger than $2$. It is known that particle chains with superquadratic potentials support the propagation of non-dispersive solitary waves \cite{FrieseckeWattis}. The existence of travelling solitary waves in Hertzian chains was first reported in \cite{Nesterenko}. Since then, 
solitary waves in Hertzian chains have been studied theoretically, numerically and experimentally~\cite{Nesterenko1,sen:2008,porter2015,chong2017}. Prior investigations of Hertzian chains have considered a wide range of behaviours, including the generation~\cite{Deng1,Lazaridi,Coste,DaraioNesterenko2006,Hinch}, propagation~\cite{Deng1,Lazaridi,Coste,DaraioNesterenko2006,Hinch}, interaction~\cite{manciu:2002,job:2005,avalos:2009,avalos:2011,avalos:2014,Deng}, and long-time dynamics \cite{avalos:2011,avalos:2014,sen:2004,Prz:2015,przedborski:2017} of solitary waves in monatomic Hertzian chain.

Hertzian chains are notable for their tunability; it is feasible to construct heterogeneous Hertzian chains in experimental settings by adding particle heterogeneities to a monatomic Hertzian chain with uniform particle properties.
Common variants of heterogenous Hertzian chains include chains with impurities~\cite{SenManciuWright,Hascoet,Martinez1}, disordered and random particle arrangements~\cite{KimMartinez,Martinez2,sen:2000,Harbola,Manjunath}, alternating masses~\cite{chong2017,Theocharis,Boechler2010,Molinari,Ponson,Porter2008,Porter2009,Hoogeboom,Herbold2009,Jayaprakash,Jayaprakash1}, quasiperiodicity~\cite{Martinez3}, and compound segments~\cite{Vergara2005,Vergara2006,NestrenkoDaraio2005,DaraioNesterenko2006_1}. Solution dynamics in heterogeneous Hertzian chains can differ significantly from that seen in monatomic Hertzian chains. 

In this study we focus on the behaviour of travelling waves in locally resonant granular chains, which are also known as woodpile chains~\cite{chong2017,Deng2021,Kim,Xu,Liu2,Liu1,Jiang,Wu,KimYang}. Woodpile chains are of particular interest not only because they can be constructed experimentally, but also because their electromagnetic counterpart, known as ``woodpile photonic crystals'', can be used to control electromagnetic waves \cite{Feigel,Liu}.

\begin{figure}[tb]
\centering
\subfloat[A physical woodpile system]{
\includegraphics{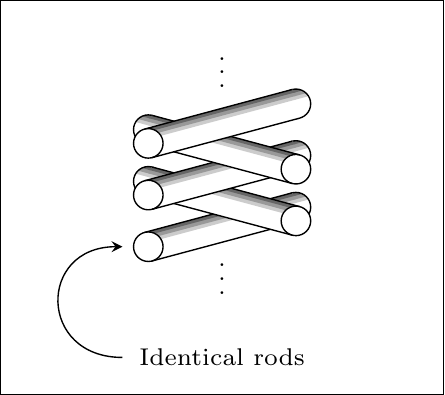}
}
\subfloat[A woodpile-chain model]{
\includegraphics{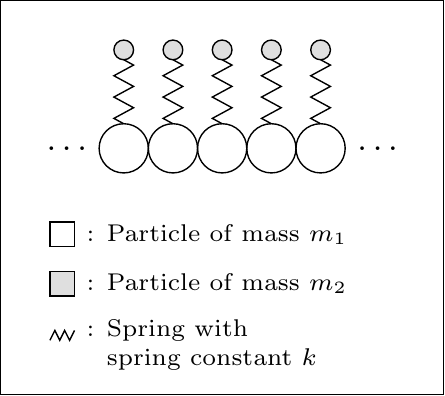}
}
\caption{The schematic in panel (a) shows the physical configuration of orthogonally stacked rods known as a ``woodpile chain''. The schematic in panel (b) shows an idealized mathematical model of the physical configuration in (a). This model consists of heavy spherical particles in physical contact via interaction potential given by~\eqref{e:potential} and light spherical particles (sometimes called ``resonators'') that are attached to each heavy particle by a spring. We analyze the model in (b) with zero precompression.
}
\label{f:DiatomicWoodpile}
\end{figure}

A woodpile chain is composed of orthogonally stacked slender rigid cylinders, illustrated in Figure~\ref{f:DiatomicWoodpile}(a). The interaction potential along the direction of the stack is given by~\eqref{e:potential}. The elastic deformation in the direction perpendicular to the stack direction is modeled by internal resonators, where the mass and coupling constant of these resonators are determined by system properties, such as the mass, shape, and material of the cylinders. For a stack of identical cylinders, each resonator has the same mass and coupling constant. The stack of identical cylinders can be modeled as a homogeneous Hertzian chain, where each chain particle is also connected to an external particle by a linear spring. We denote the mass of particles in the monatomic Hertzian chain by $m_1$, the mass of external particles by $m_2$ and the spring constant by $k$. The configuration of this idealized model of the woodpile chain is shown in Figure~\ref{f:DiatomicWoodpile}(b).

In earlier studies, various properties of the woodpile chains were investigated, such as their sound  absorbing properties~\cite{Jiang} and their frequency band structure~\cite{Wu,KimYang}. These studies were focused on the behaviour of woodpile chains in the linear regime, which is obtained by expanding the interaction potential~\eqref{e:potential} and keeping only up to the leading term for large $\Delta$. Woodpile chains in the nonlinear regime behave in a fashion which is distinct from either a monatomic Hertizan chain or a linear system. The existence of discrete breathers for  woodpile chains in the weakly and strongly nonlinear regime was reported in~\cite{Liu1,Liu2}. Most significantly for the present study, travelling-wave solutions in woodpile chains in the weakly~\cite{Deng2021} and strongly~\cite{Kim,Xu} nonlinear regime have been shown to be nanoptera.

\begin{figure}[b!]
\centering
\subfloat[A solitary wave]{
\includegraphics{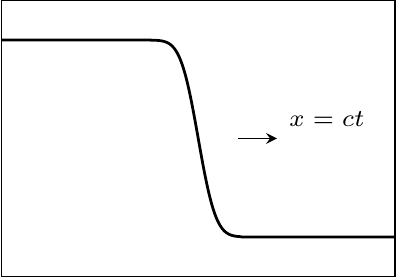}
}
\subfloat[A one-sided nanopteron]{
\includegraphics{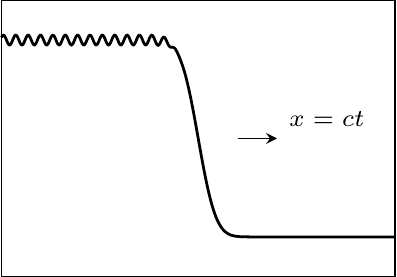}
}
\subfloat[A two-sided nanopteron]{
\includegraphics{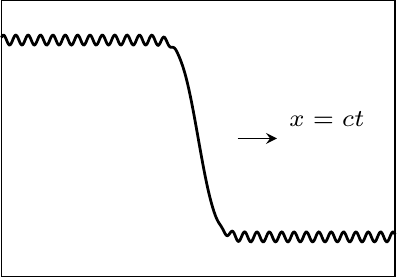}
}
\caption{Comparison of the profiles of (a) a standard solitary wave, (b) a one-sided nanopteron, and (c) a two-sided nanopteron that each propagates at speed $c$. The solitary wave is localized spatially, whereas the nanoptera have non-decaying oscillatory tails on (b) one side or (c) both sides of the wave front. The waves in panels (a) and (c) propagate without decaying, but the wave in panel (b) cannot propagate indefinitely. In (b), the one-sided oscillation draws energy from the wave front, leading to the eventual decay of the wave front. As this decay occurs over a long time scale, the one-sided nanopteron in (b) is said to be ``metastable''.
}
\label{f:nanoptera}
\end{figure}

\subsection{Nanoptera}
A typical travelling-wave solution in a woodpile chain is not a localized solitary wave~\cite{Kim,Xu}. Instead, these solutions take the form of nanoptera~\cite{Boyd}. A nanopteron is composed of a solitary wave and tailing oscillations on one or both sides of the central wave that extend indefinitely without decaying. The amplitude of the tailing oscillations is usually exponentially small in some asymptotic parameter of the problem. In Figure \ref{f:nanoptera}, we illustrate examples of a solitary wave, a one-sided nanopteron and a two-sided nanopteron respectively. While a true solitary wave is exponentially localized in space, a nanopteron is localized only up to algebraic orders in the small parameter.

Travelling-wave solutions in various diatomic chains, such as diatomic Toda~\cite{Vainchtein,Okada,Tabata,Lustri},  FPUT~\cite{Vainchtein,Faver,Hoffman,Lustri1} and Hertzian chains~\cite{Deng2021}, also take the form of nanoptera. Nanoptera solutions have been identified in chains with an on-site nonlinear potential, where each particle in the chain couples linearly to its neighbors~\cite{Iooss}. A rigorous proof of the existence of nanoptera solutions in the diatomic FPUT chain was shown in~\cite{Faver,Hoffman}. In~\cite{Vainchtein}, singular multiscale asymptotic analysis is applied to investigate nanoptera solutions in diatomic chains with small mass ratio. More recently,~\cite{Faver2} explored how nanopteron solutions for chains with small mass ratio are related to travelling-wave solutions for chains where the mass ratio is not small.

It has been shown that for some singularly-perturbed chains, an appropriate choice of system parameters causes the oscillations in the tail to vanish entirely, producing localized solitary waves. These parameter choices satisfy a condition known as an anti-resonance condition. Anti-resonance conditions have been studied in different types of diatomic chains, including diatomic Hertzian~\cite{Jayaprakash,KimChaunsali2015,Potekin2013,Deng,Manjunath2014}, Toda~\cite{Vainchtein,Tabata,Lustri} and FPUT chains~\cite{Lustri}, and in a discrete nonlinear Schr\"{o}dinger equation~\cite{Alfimov}. It is particularly significant in the context of the present analysis to note that anti-resonance conditions have been identified in woodpile chains with zero precompression by studying the Fourier transform of the system~\cite{Xu}. Nanoptera do not decay infinitely far from the solitary wave, and therefore the Fourier transform does not exist for them; hence, the existence condition for the Fourier transform implies that the tailing oscillations disappear, and is therefore equivalent to an anti-resonance condition.

More recently, studies in~\cite{Lustri,Lustri1,Deng2021} used exponential asymptotics to calculate asymptotic expressions for nanopteron solutions in a number of diatomic chains, as well as woodpile chains with strong precompression. These studies demonstrate that the tailing oscillations following the central leading wave can be explained by the Stokes phenomenon, which refers to behaviour that is switched on when curves known as ``Stokes curves" are crossed in the complex plane. Travelling waves in these diatomic chains were shown to contain two Stokes curves, generating two sets of oscillations with same amplitude but different phases. These two oscillations vanish entirely when they are precisely out of phase. In contrast, travelling waves in strongly precompressed woodpile chains contain a single Stokes curve, which does not vanish for any parameter choice. Therefore woodpile chains with strong precompression do not possess any anti-resonance condition.

In Figure~\ref{f:nanoptera} we illustrate two types of nanoptera, containing oscillations on one or both sides of the central leading wave. The analysis presented in~\cite{Hunter1988} showed that symmetric nanoptera, for which the amplitudes of the oscillations on both sides are the same, exist for the fifth-order Korteweg-de Vries (KdV) equation. Later,~\cite{Boyd1} explicitly constructed symmetric nanoptera for the fifth-order KdV equation, but numerical attempts to obtain asymmetric one-sided nanoptera to arbitrary order were unsuccessful. It was later shown that one-sided nanoptera for the fifth-order KdV equation cannot propagate permanently without decaying~\cite{Benilov,Grimshaw}, as the one-sided oscillation absorbs energy from the central wave leading to its eventual decay. For any system, which is energy-conserving, such as granular chains, we can make a similar argument to show that one-sided nanoptera cannot propagate indefinitely.

One-sided nanoptera in diatomic FPUT chains were investigated in~\cite{Giardetti}. The study showed that for small mass ratio, nanoptera with small amplitude can travel without any evident decay for an extended duration. Nonetheless, the one-sided tailing oscillation must constantly draw energy from the central wave~\cite{Giardetti,Jayaprakash,Vainchtein}. The total energy of a diatomic chain without driving or damping force must be conserved, it is therefore impossible for one-sided nanoptera to propagate indefinitely without changing shape. In~\cite{Giardetti}, the authors conjectured that the decay of one-sided nanoptera in the diatomic FPUT lattice occurs on a time scale that is exponentially large in the limit that the mass ratio becomes small. Solutions which appear to be stable over a time scale that is large comparing to any inverse power of the small asymptotic parameter, are known as ``metastable" or ``quasistable" solutions. The nanoptera considered in the present study are all metastable. 

\subsection{Paper outline}

We compute the behaviour of nanopteron solutions in woodpile chains with zero precompression in the asymptotic limit that $m_2/m_1\to0$. In this limit, the woodpile chain is singularly perturbed around a monoatomic granular chain. The study in~\cite{Deng2021} investigated this limit for woodpile chains with strong precompression, which is a weakly nonlinear regime. We will compare the behaviour of traveling waves in the two regimes. We will then demonstrate how the Stokes structure, and therefore the system behaviour, changes as the system varies between the two regimes. 

We apply an exponential asymptotic method, which was developed in~\cite{Chapman,Daalhuis} and was used in \cite{Lustri,Lustri1,Deng2021} to study travelling waves in diatomic Toda, FPUT and Hertzian chains with small mass ratios. In a typical asymptotic power series analysis, the solution is expanded as a power series in some small parameter. The terms of this series are then computed using a recursion relation generated by the asymptotic matching of series terms. This process can never capture exponentially small behaviour, which is smaller than any algebraic series term in the asymptotic limit. Exponential asymptotic methods, such as that used  in the present paper, are capable of describing asymptotic behaviour on this exponentially small scale.

To apply exponential asymptotic methods to study the Stokes phenomenon, we require an expression for the leading-order solitary wave solution. Analytically continuing this expression typically reveals singular points in the complex plane, which are the end points of Stokes curves. In previous studies, the leading-order behaviour can be calculated exactly~\cite{Lustri} or approximated analytically using a weakly nonlinear KdV equation~\cite{Deng2021,Lustri1}. For woodpile chains with zero precompression, it is not possible to obtain an exact solution, or even a weakly nonlinear approximation, at leading order.

This is a significant obstacle to overcome, as it is impossible to write down analytic solutions for many nonlinear systems. Numerical leading-order solutions have been obtained in exponential asymptotic studies of parasitic gravity-capillary waves~\cite{Shelton} and thin film rupture~\cite{Chapman2013}. These studies used purely numerical methods to approximate the leading-order singularity locations. Instead, we adopt the hybrid numerical-analytic method from~\cite{sen2001}, which uses numerical data to fit coefficients of an analytic leading-order approximation. We use this approximated solution as the basis for the exponential asymptotic method.

Using this method, we find that the leading-order solution possesses two important Stokes curves, which generate two oscillatory contributions in the wake of the leading wave. This differs significantly from the weakly nonlinear regime, in which only a single oscillatory contribution exists. We obtain simple asymptotic expressions for the exponentially small, constant-amplitude oscillations, and determine special mass ratios at which the two oscillatory contributions cancel precisely and the travelling-wave solution becomes a localized solitary wave.

By introducing precompression into the woodpile chain model, we show that as the precompression becomes stronger, the two Stokes curves seen in the fully nonlinear regime approach each other. We then show that there exists a critical value for the precompression at which the two Stokes curves coalesce into a single curve. Above this critical value, nanopteron solutions contain a single oscillatory contribution in the wake that never vanishes. This is consistent with \cite{Deng2021}, which found that anti-resonance conditions do not exist for strongly precompressed chains. 

Our paper proceeds as follows. In Section~\ref{s:exponential}, we introduce the exponential asymptotic method that we employ for our analysis. This method is essentially identical to that used in~\cite{Lustri,Lustri1,Deng2021}. In Sections~\ref{s:woodLO}--\ref{s:woodstokes}, we use this exponential asymptotic method to obtain an asymptotic approximation for nanopteron solutions in a singularly perturbed woodpile chain with zero precompression, in which the leading-order solution is approximated using the hybrid numerical-analytic technique of~\cite{sen2001}. In Section~\ref{s:woodnum}, we compare the asymptotic results with numerical computations. In Section~\ref{s:transition}, we investigate the behaviour of Stokes curves in the travelling wave solution as precompression increases, approaching the weakly nonlinear regime associated with strong precompression.  We present our conclusions and discuss the results further in Section~\ref{s:conclusion}.

%%%%%%%%%%%%%%%%%%%%%%%%%%%%%%%%%%%%%%%%%%%%%%%%%%%%%%%%%%%%%%%%%%%%%%%%%%%%%%%%%%%%%%%%%
\section{Exponential asymptotics}
\label{s:exponential}
The methodology used in this study is very similar to that of~\cite{Lustri,Lustri1,Deng2021}. The explanation of the methodology is therefore similar to the explanation contained in these previous studies.

Our goal is to calculate the behaviour of exponentially small oscillations in the wake of a leading-order solitary wave where the mass ratio $\eta^2=m_2/m_1$ is small. This system is singularly perturbed in the limit $\eta\to0$. Using classical perturbation methods, we can expand the solution as a power series in $\eta$. The exponentially small oscillations, however, are smaller than any term in the power series in the limit $\eta\to0$. Therefore we cannot determine the asymptotic behaviour of these oscillations using classical power series methods. We must instead use an exponential asymptotic approach which is able to calculate exponentially small asymptotic effects.

We consider a singularly perturbed differential equation of the following form
\begin{equation}
	F(x,g(x),g'(x),g''(x), \ldots;\eta)=0\,,
\label{e:intro_equation}
\end{equation}
where $\eta$ is a small parameter. We first determine the leading-order solution, which is obtained by setting $\eta=0$ in~\eqref{e:intro_equation}. We then analytically continue the leading-order solution into the complex plane. This solution typically contains a set of singularities in the complex plane, and these singularities are the end points of Stokes curves~\cite{Stokes}. As Stokes curves are crossed, there is a rapid change in the amplitude of the exponentially small contribution, known as ``Stokes switching''. One-sided nanoptera appear if the exponentially small contribution is absent on one side of the Stokes curve and ``switches on'' as the Stokes curve is crossed.

We subsequently expand the solution about the leading-order behaviour as a power series in the parameter $\eta$ to obtain an asymptotic solution to~\eqref{e:intro_equation}.
In general, the asymptotic power series solution to a singularly perturbed problem is divergent~\cite{Dingle}, but we can truncate the power series to obtain an approximation of the solution~\cite{Boyd2005}. When the truncation point is chosen so that the difference between the exact and approximated solutions is minimized, the approximation error is exponentially small in the asymptotic limit~\cite{Boyd1999}. The truncation point that minimizes the approximation error is known as ``optimal truncation point'', and we denote this value as $N_{\mathrm{opt}}$. The solution can therefore be expressed as a sum of an optimally truncated power series and an exponentially small error term. By substituting this sum into~\eqref{e:intro_equation}, we can obtain an equation governing the exponentially small term.

This idea was developed in~\cite{Berry1988,Berry}, and applied to determine the Stokes switching behaviour in several important special functions. Subsequently,~\cite{BerryHowls1990} established the hyperasymptotics techniques to further reduce the exponentially small error generated by truncating the series. See~\cite{Berry1991} for a summary and discussion of results in~\cite{Berry,BerryHowls1990}.

In the present study, we apply an exponential asymptotic method that was developed in~\cite{Chapman,Daalhuis}. We express the solution $g$ of governing equation~\eqref{e:intro_equation} as an asymptotic power series
\begin{equation}\label{e:series_intro}
	g \sim \sum_{j=0}^\infty \eta^{r j}g_j \quad \mathrm{as} \quad \eta \rightarrow 0\,,
\end{equation}
where $r$ is the number of times that $g_{j-1}$ must be differentiated to obtain $g_j$.

By substituting the series \eqref{e:series_intro} into the governing equation~\eqref{e:intro_equation} and matching terms at the same order of $\eta$, we obtain a recursion relation for $g_j$. In a singularly perturbed problem, applying the recursion relation to obtain $g_j$ requires differentiating earlier terms in the power series. For series with terms containing singular points, the repeated differentiation guarantees terms in the series diverge in a predictable fashion. This form of divergence is known as ``factorial-over-power divergence"~\cite{Dingle}. Behaviour with this form dominates the series terms for large $j$.

To capture this divergence, it is necessary to obtain an ansatz for the behaviour of $g_j$ in the limit that $j \rightarrow \infty$. The terms $g_j$ with large $j$ are also known as ``late-order terms''. In ~\cite{Chapman}, the authors proposed applying a late-order ansatz to approximate the form of the late-order terms,
\begin{equation}
	g_j\sim\frac{G\Gamma(r j+\gamma)}{\chi^{r j+\gamma}}\quad \mathrm{as} \quad j\to\infty\,,
\label{e:lateorder_intro}
\end{equation}
where the parameter $\gamma$ is constant and $G$ and $\chi$ are functions of any independent variables but are independent of $j$. The function $\chi$ is known as the ``singulant". To ensure that the late-order terms $g_j$ are singular at the same locations as the leading-order solution, the singulant is equal to $0$ at each singularity of the leading-order solution $g_0$. With $j$ increases the singularity strength of the late-order terms also grows. By combining~\eqref{e:intro_equation}~\eqref{e:series_intro} and~\eqref{e:lateorder_intro}, and matching orders of $\eta$, we can obtain functional forms of $\chi$ and $G$. We can determine $\gamma$ by requiring that the late-order behaviour is consistent with the local behaviour of the leading-order solution in the neighborhood of singular points. As shown in~\cite{BerryHowls1990},  Stokes curves follow curves on which $\chi$ is real and positive.

A heuristic for determining the optimal truncation point is given in~\cite{Boyd1999}. This heuristic requires truncating the series at the value of $N_{\mathrm{opt}}$ for which the term $\eta^{N_{\mathrm{opt}}} g_{N_{\mathrm{opt}}}$ has the smallest magnitude. As $N_{\mathrm{opt}}$ is typically large, we can apply the late-order ansatz~\eqref{e:lateorder_intro}, which is valid for large $j$ to determine the optimal truncation point.

We then write the solution in the following form
\begin{equation}
	g = \sum_{j=0}^{N_{\mathrm{opt}}-1} \eta^{r j}g_j+g_{\exp}\,,
\label{e:series_intro_1}
\end{equation}
where $g_{\exp}$ denotes the exponentially small error term and $N_{\mathrm{opt}}$ is the optimal truncation point.

By substituting the truncated series expression \eqref{e:series_intro_1} into the differential equation \eqref{e:intro_equation}, we can obtain an equation for the exponentially small remainder term~\cite{Daalhuis}. Away from the Stokes curves, we can find this remainder by applying the Liouville-Green (or WKB) method \cite{hinch1991}. In the neighborhood of the Stokes curves, we apply the following exponential ansatz for the exponentially small asymptotic term $g_{\exp}$
\begin{equation}
	g_{\exp}\sim\mathcal{S}G\e^{-\chi/\eta} \quad \mathrm{as} \quad \eta \rightarrow 0\,,
\label{e:stokes_intro}
\end{equation}
where $\mathcal{S}$ is the Stokes multiplier. Away from the Stokes curve, the Stokes multiplier takes constant value, and~\eqref{e:stokes_intro} reduces to the standard Liouville-Green ansatz. In the transition region of width $\mathcal{O}(\sqrt{\eta})$ as $\eta\to0$ around the Stokes curves, the Stokes multiplier undergoes a rapid change, known as Stokes switching. Substituting the exponential ansatz~\eqref{e:stokes_intro} into the governing equation for $g_{\exp}$, we can obtain the exponentially small contributions that appear as the Stokes curves are crossed.  This contribution can never be obtained using classical asymptotic power series.

Using this method requires to obtaining an explicit expression only for the leading-order solution in~\eqref{e:series_intro}. This makes it convenient to apply the exponential asymptotic method to nonlinear problems, where computing series terms beyond a leading-order expression can be challenging. 
For more details on exponential asymptotics and their applications to nonlocal solitary waves see~\cite{Boyd1998,Boyd1999}, for examples of other studies of exponential asymptotics see~\cite{Berry,Berry1991}, and for more details on the particular methodology that we apply in the present paper see~\cite{Chapman,Daalhuis}.

%%%%%%%%%%%%%%%%%%%%%%%%%%%%%%%%%%%%%%%%%%%%%%%%%%%%%%%%%%%%%%%%%%%%%%%%%%%%%%%%%%%%%%%%%
\section{Woodpile chain with zero precompression}

\label{s:woodpile}
We model the woodpile chain from Figure~\ref{f:DiatomicWoodpile}(a) as a singularly perturbed Hertzian chain, where each particle in a homogeneous Hertzian chain with mass $m_1$ is connected to an outside particle with mass $m_2$ by a harmonic spring with elastic constant $k$. This idealized model is shown in Figure~\ref{f:DiatomicWoodpile}(b). In all subsequent discussion, the term ``woodpile chain'' refers to this idealized model. The governing equations of a woodpile chain are
\begin{align}\label{1:gov0a}
	m_1 \ddot{u}(n,t) &= [\Delta  + u(n-1,t) - u(n,t)]_+^\alpha - [\Delta  + u(n,t) - u(n+1,t)]_+^\alpha - k [u(n,t) - v(n,t)]\,,\\
	m_2 \ddot{v}(n,t) &= k [u(n,t) - v(n,t)]\,,
	\label{1:gov0b}
\end{align}
where $\Delta$ is the precompression, and $u(n,t)$ and $v(n,t)$, respectively, denote the displacement of the $n$-th particle of mass $m_1$ and $m_2$ at time $t$. In this section, we consider a chain with zero precompression, or $\Delta=0$. The analysis in this section is valid for any choice of $\alpha$ where the system supports the propagation of solitary waves, which was shown in~\cite{FrieseckeWattis} to include only chains with superquadratic potential, or $\alpha > 1$. Setting $\alpha=3/2$ gives the equations governing a woodpile chain with classical Hertzian interaction potential. The subscript $+$ indicates that we evaluate the bracketed term only if its argument is positive; it is equal to zero otherwise. That is, the interaction effects only happen between particles that are in physical contact with each other.

Next we scale the system \eqref{1:gov0a}--\eqref{1:gov0b} with $u= m_1^2 \hat{u}$ and $v= m_1^2 \hat{v}$. Introducing $\hat{k} = k/m_1$ and $\eta^2 = m_2/m_1$, we write~\eqref{1:gov0a}--\eqref{1:gov0b} as the following scaled governing equations
\begin{align}\label{1:gov1}
	\ddot{\hat{u}}(n,t) &= [\hat{u}(n-1,t) - \hat{u}(n,t)]_+^\alpha - [\hat{u}(n,t) - \hat{u}(n+1,t)]_+^\alpha - \hat{k} [\hat{u}(n,t) - \hat{v}(n,t)]\,,\\
\eta^2 \ddot{\hat{v}}(n,t) &= \hat{k} [\hat{u}(n,t) - \hat{v}(n,t)]\,.
\label{1:gov2}
\end{align}
In the following sections, we perform our analysis on the scaled system \eqref{1:gov1}--\eqref{1:gov2}. For simplicity of notation, we omit the hats in subsequent analysis. We consider the asymptotic behaviour of the system for small values of the mass ratio, or $0<\eta\ll1$.

%%%%

\subsection{Leading-order solution}\label{s:woodLO}

In the limit $\eta \rightarrow 0$ we write $u(x,t)$ and $v(x,t)$ as asymptotic power series in $\eta^2$
\begin{align}
	u(n,t)\sim\sum_{j=0}^\infty \eta^{2j}u_j(n,t)\,, \quad v(n,t)\sim\sum_{j=0}^\infty \eta^{2j}v_j(n,t)\,.
\label{1:asympseries}
\end{align}

To construct a nanopteron solution of \eqref{1:gov1}--\eqref{1:gov2}, we first identify the behaviour of the leading-order solitary wave.
Substituting the series expression~\eqref{1:asympseries} into~\eqref{1:gov2} and matching at the leading order in the limit $\eta\to0$, we obtain
\begin{align}
	u_0(n,t) = v_0(n,t)\,.
\label{1:leading}
\end{align}
Inserting~\eqref{1:leading} into
\eqref{1:gov1} gives
\begin{align}
	\ddot u_0(n,t)=\left[u_0({n-1},t)-u_0({n},t)\right]_+^\alpha-\left[u_0({n},t)-u_0({n+1},t)\right]_+^\alpha\,.
\label{1:woodpile12}
\end{align}

In the strongly precompressed version of this problem~\cite{Deng2021}, we can take the long wave limit of the system to obtain the KdV equation. The leading-order solitary wave is therefore approximated by the well-known KdV soliton. In the zero precompression regime,~\eqref{1:woodpile12} does not have a linear limit, and cannot be transformed in this way. We instead apply a hybrid numerical--analytic method introduced in~\cite{sen2001} to approximate the leading-order solitary wave for~\eqref{1:woodpile12}.

In~\cite{sen2001}, the authors note that all solitary waves in the zero precompression regime have identical width, and that there is a simple scaling relation between the solitary wave amplitude and the solitary wave velocity. Motivated by this observation, solitary waves in~\eqref{1:woodpile12} are assumed to take the form
\begin{align}
u_0(\xi) \approx \frac{A}{2}[1-\tanh(f(\xi)/2)],\quad f(\xi)=\sum_{n=0}^N C_{2n+1}\xi^{2n+1}, \quad \xi=n-c_A t,
\label{e:senmanciu}
\end{align}
where $A$ is the amplitude of the solitary wave, $c_A$ is the velocity of the solitary wave with amplitude $A$, $N$ is a positive integer, and
$\xi$ defines the co-moving frame. This approximation is valid for a solitary wave profile for which $u_0(\xi) \to A$ as $\xi \to -\infty$ and $u_0(\xi)\to 0$ as $\xi \to \infty$. The scaling relation between $c_A$ and $A$ is given by
\begin{align}
\label{e:scaling}
c_A = C_0(\alpha)^{1/2}(A/2)^{(\alpha-1)/2},
\end{align}
where $C_0$ depends only on the exponent $\alpha$ and can be determined numerically~\cite{sen2001}. This form of solitary wave solutions is distinct from that in the weakly nonlinear regime, in which the width of the wave depends on the amplitude parameter.

\begin{figure}[tb]
\centering

\subfloat[$f(\xi)$ approximated using $N = 2$, where $C_5 > 0$]{
%\begin{tikzpicture}
%[x =0.2cm,y =3cm]
%\node at (-10,0) [left] {\scriptsize{$0$}};
%\node at (-10,1) [left] {\scriptsize{$1$}};
%\node at (10,-0.05) [below] {\scriptsize{$10$}};
%\node at (-10,-0.05) [below] {\scriptsize{$-10$}};
%\node at (0,-0.05) [below] {\scriptsize{$0$}};
%
%\draw[gray!50,line width=1mm,dotted] plot[smooth] file {coefficient_exact.txt};
%\draw[black] plot[smooth,line width=0.75mm] file {coefficient_positive.txt};
%
%\draw (1,0.8) -- (3,0.8) node[right] {\scriptsize{Approx.}};
%\draw[gray!50,line width=1mm,dotted] (1,0.65) -- (3,0.65);
%\node at (3,0.65) [right] {\scriptsize{Exact}};
%
%\draw node at (-11.5,0.5) {\scriptsize{$u_0$}};
%\draw node at (0,-0.25) {\scriptsize{$\xi$}};
%\draw (-10,-0.05) -- (-10,1.05) -- (10,1.05) --(10,-0.05) -- cycle;
%\end{tikzpicture}
\includegraphics{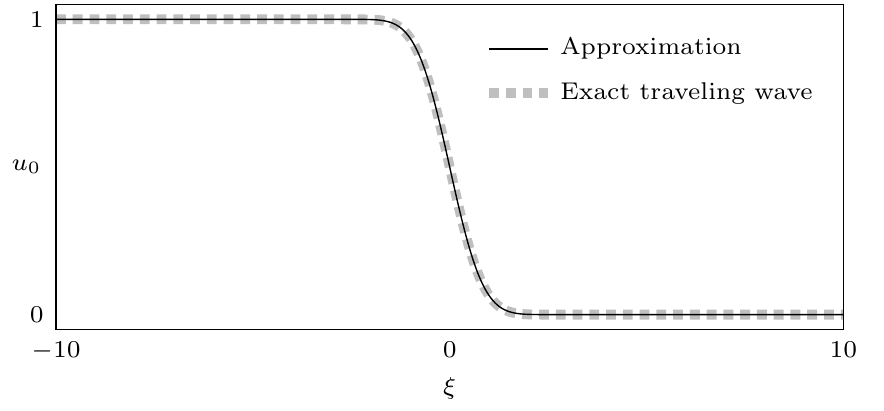}
}%\hspace{2mm}
\subfloat[$f(\xi)$ approximated using $N = 3$, where $C_7 < 0$]{
%\begin{tikzpicture}
%[x =0.2cm,y =3cm]
%\node at (-10,0) [left] {\scriptsize{$0$}};
%\node at (-10,1) [left] {\scriptsize{$1$}};
%\node at (10,-0.05) [below] {\scriptsize{$10$}};
%\node at (-10,-0.05) [below] {\scriptsize{$-10$}};
%\node at (0,-0.05) [below] {\scriptsize{$0$}};
%
%\draw[gray!50,line width=1mm,dotted] plot[smooth] file {coefficient_exact.txt};
%\draw[black] plot[smooth,line width=0.75mm] file {coefficient_negative_1.txt};
%
%\draw node at (-11.5,0.5) {\scriptsize{$u_0$}};
%\draw node at (0,-0.25) {\scriptsize{$\xi$}};
%\draw (-10,-0.05) -- (-10,1.05) -- (10,1.05) --(10,-0.05) -- cycle;
%\end{tikzpicture}
\includegraphics{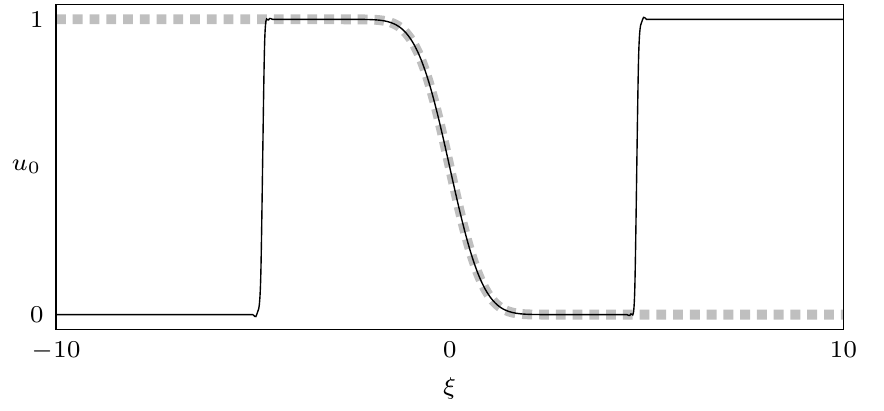}
}
\caption{Leading-order travelling wave $u_0(\xi)$ for $\alpha=1.5$ and $A=1$. Figure (a) shows the solution where $f(\xi)$ is approximated using the expression from~\eqref{e:senmanciu} with $N=2$, including coefficients up to $C_5$, where $C_5 = 0.0061$. Figure (b) shows the solution approximated with $N=3$, including coefficients up to $C_7$, where $C_7=-0.0010$. In each case, the gray dotted curve represents the numerically computed exact solution. It is clear that the solution including $C_7$ becomes inaccurate away from $\xi = 0$. 
}
\label{f:fitting_coefficient}
\end{figure}

The method in~\cite{sen2001} requires differentiating~\eqref{e:senmanciu} $2n+1$ times with respect to $\xi$. Evaluating the result at $\xi=0$, we obtain a polynomial containing only $C_{2j+1}$ terms with $j\leq n$. 
The numerical solitary wave, generated by applying a velocity impulse to a simulated chain, is also differentiated $2n+1$ times and evaluated at $\xi = 0$.
By equating the analytically and numerically differentiated expressions, we then determine coefficients $C_{2n+1}$ recursively.
The values of the coefficients are reported in Table~\ref{t:1}. For $\alpha=1.2,1.5,2,3$ the values of $C_{2n+1}$ are taken directly from~\cite{sen2001}; for the other values of $\alpha$ the values of $C_{2n+1}$  are computed following the method in~\cite{sen2001}. It is worth noting that in the zero precompression regime the coefficients $C_{2n+1}$, which are determined by $\alpha$, do not depend on the amplitude of the solitary wave.

This process cannot be extended indefinitely. In each of the examples considered here, this method produces a negative value for $C_7$. As illustrated in Figure~\ref{f:fitting_coefficient}, when including negative coefficient $C_7$ the approximated profile behaves correctly in the vicinity of $\xi = 0$, but the wave profile cannot be correct for large $|\xi|$. In this case, $u_0(\xi) \to A$  as $\xi \to \infty$ and $u_0(\xi) \to 0$ as $\xi \to -\infty$, despite the derivative being negative at $\xi=0$. We therefore only consider approximations here using coefficients up to $C_5$, as in~\cite{sen2001}.

We note that while alternative methods for approximating the leading-order solitary wave to arbitrary accuracy are viable, such as rational approximation methods~\cite{pade,Nakatsukasa}, the approximation obtained in~\eqref{e:senmanciu} is sufficiently accurate for the purposes of the present study.

\begin{table}[htbp]
\caption{Coefficients $C_n$ for different values of $\alpha$.}
\label{t:1}
\centering
\begin{tabular}{|c |c c c c|} \hline
 $\alpha$ & $C_0$  &  $C_1$ &$C_3$ &  $C_5$ \\ \hline
1.2       & 0.8710  &1.6437      &0.0822   &0.0003\\ \hline
1.5       & 0.8585 & 2.3954  & 0.2685& 0.0061 \\ \hline
1.75      & 0.8913 & 2.7653  & 0.4373 & 0.0229 \\ \hline
2         & 0.9445  & 3.0168  & 0.5971 & 0.0376 \\ \hline
2.5       & 1.1027  & 3.3545  & 0.9418 & 0.0622 \\ \hline
3         & 1.33237  &3.56461  & 1.3314& 0.0676 \\ \hline
3.5       & 1.6532 &3.7101 &1.7763  & 0.1019\\ \hline
\end{tabular}
\end{table}

When analytically continued such that $\xi \in \mathbb{C}$, the leading-order behaviour in \eqref{e:senmanciu} contains singularities at values of $\xi$ satisfying $f(\xi) = (2M+1)\pi\i$, for $M \in \mathbb{Z}$. These singularities will determine the behaviour of the late-order series terms in \eqref{1:asympseries}. In practice, the late-order behaviour is dominated by contributions from singularities closest to the real axis, which satisfy $f(\xi) = \pm\pi\i$. Solving this expression reveals that there are four singularities -- one in each quadrant of the complex plane -- that contribute to the late-order term behaviour, as they are equally close to the real axis. We denote the singularity with positive real and imaginary part as $\xi_s$. The remaining singularities are its complex conjugate $\xi^*_s$, and singularities located at $-\xi_s$ and $-\xi^*_s$. An example of this configuration is shown in Figure~\ref{f:stokes}. 

To make our discussion more general, we will consider the late-order terms caused by a singularity located at some $\xi = \xi_s$, and consider particular values for $\alpha$ and $\xi_s$ at the conclusion of the general analysis. Expanding~\eqref{e:senmanciu} near $\xi = \xi_s$ shows that
\begin{align}\label{1:u0loc}
	u_0(\xi_s) \sim -\frac{\bar{A}(\xi_s)}{\xi - \xi_s} \quad \mathrm{as} \quad \xi \rightarrow \xi_s,
\end{align}
where
\begin{align}
\label{e:barA}
\bar{A}(\xi_s)=\frac{A}{C_1+3C_3\xi_s^2+5C_5\xi_s^4},
\end{align}
and that $v_0$ has the same behaviour as $u_0$ in the neighborhood of the singularity.

%%%%%%%%%%%%%%%%%%%%%%%%%%%%%%%%%%%%%%%%%%%%%%%%%%%%%%
\subsection{Late-order terms}
\label{s:late order}

Writing the governing equations \eqref{1:gov1}--\eqref{1:gov2} in terms of $\xi$ and matching at each order of $\eta$ gives
\begin{align}\label{2:govser1}
	\nonumber c_A^2 u_j''(\xi) &= \tfrac{3}{2}[u_j(\xi-1) - u_j(\xi))(u_0(\xi-1) - u_0(\xi)]^{\alpha-1} \\&\quad- \tfrac{3}{2}[u_j(\xi) - u_j(\xi+1)][u_0(\xi) - u_0(\xi+1)]^{\alpha-1} - k [u_j(\xi) - v_j(\xi)] + \ldots\,,\\
	c_A^2 v_{j-1}''(\xi) &= k [u_j(\xi) - v_j(\xi)]\,,
	\label{2:govser2}
\end{align}
where we only retain terms containing $u_j$, $v_j$, and derivatives of $u_j$ and $v_{j-1}$.

Terms which are products including $u_{j-k}$ with $k>1$ are omitted. It is apparent from the general form of the factorial-over-power ansatz \eqref{e:lateorder_intro} that the omitted terms are subdominant comparing to those retained in the limit $j\to\infty$.

In principle, by applying~\eqref{2:govser1} and~\eqref{2:govser2} recursively we can obtain terms in the series~\eqref{1:asympseries} up to arbitrary order. In practice this process can be challenging or intractable, as we need to solve both a differential--difference equation~\eqref{2:govser1} and an algebraic equation~\eqref{2:govser2} at each order. Furthermore, obtaining terms in the series~\eqref{1:asympseries} up to arbitrary order does not capture the behaviour of the exponentially small oscillations as $\xi\to-\infty$, as the oscillations are smaller than any term in~\eqref{1:asympseries} as $\eta\to0$. 

Instead, the next stage of the analysis is to find an asymptotic form of the late-order terms. We pose a late-order ansatz that consists of sums of terms with the form
\begin{equation}\label{2:ansatz}
	u_j \sim \frac{U(\xi)\Gamma(2j + \beta_1)}{\chi(\xi)^{2j + \beta_1}}\,, \quad v_j \sim \frac{V(\xi)\Gamma(2j + \beta_2)}{\chi(\xi)^{2j + \beta_2}} \quad \mathrm{as} \quad j \rightarrow \infty\,.
\end{equation}
We set $\chi=0$ at $\xi = \xi_s$ so that late-order terms are singular at the same locations as the leading-order solution. 

The late-order terms~\eqref{2:ansatz} diverge in a factorial-over-power fashion as $j\to\infty$, confirming that $u_j\gg u_{j-k}$ and $v_j\gg v_{j-k}$ for $k > 0$ as $j\to\infty$. By applying the late-order ansatz~\eqref{2:ansatz} into~\eqref{2:govser1}, we find that only $\beta_1 + 2 = \beta_2$ can produce a nontrivial asymptotic balance,. This implies that $u_j = \mathcal{O}(v_{j-1})$ as $j \rightarrow \infty$, and therefore that $v_j \gg u_j$ as $j\to\infty$.

Applying the late-order ansatz \eqref{2:ansatz} to~\eqref{2:govser2} gives
\begin{align}\nonumber
	 \frac{c_A^2 (\chi'(\xi))^2 V(\xi)\Gamma(2j + \beta_2)}{\chi(\xi)^{2j + \beta_2}}& - \frac{2c_A^2 \chi'(\xi) V'(\xi)\Gamma(2j + \beta_2-1)}{\chi(\xi)^{2j + \beta_2-1}}\\&
 	- \frac{c_A^2 \chi''(\xi) V(\xi) \Gamma(2j + \beta_2 - 1)}{\chi(\xi)^{2j + \beta_2 - 1}} + \cdots = -\frac{k V(\xi)\Gamma(2j + \beta_2)}{\chi(\xi)^{2j + \beta_2}} + \cdots\,,
\end{align}
where the omitted terms are no larger than $\mathcal{O}(v_{j-1})$ in the $j \rightarrow \infty$ limit.

Matching terms at $\mathcal{O}(v_j)$ in the $j \rightarrow \infty$ limit, we obtain the singulant equation 
\begin{equation}
c_A^2(\chi'(\xi))^2 = -k,
\label{e:singulant_woodpile}
\end{equation}
implying $\chi'(\xi) = \pm \i\sqrt{k}/c_A$. We integrate to obtain
\begin{equation}\label{2:singulant}
	\chi(\xi) = \pm\frac{ \i \sqrt{k}(\xi-\xi_s)}{c_A}\,.
\end{equation}
Note that the expression for $\chi$~\eqref{2:singulant} is the same as that in strongly precompressed woodpile chains \cite{Deng2021}.
The form of $\chi$ determines the locations of the Stokes curves. Stokes curves only occur where $\mathrm{Im}(\chi) = 0$ and $\mathrm{Re}(\chi) > 0$. This corresponds to the positive sign choice for $\xi_s$ and $-\xi_s$, and the negative sign choice for $\xi_s^*$ and $-\xi_s^*$. 
We do not consider late-order terms associated with the remaining sign choices, as they do not generate Stokes curves. We restrict our attention to singularity at $\xi_s$, and will state the equivalent results for the other singularities at the conclusion of the subsequent analysis.

Matching terms at $\mathcal{O}(v'_{j-1})$, we obtain the prefactor equation $2 V'(\xi)\chi'(\xi) = 0$. This implies that $V$ is constant, with a value that depends on the choice of singularity. We denote the constant prefactor that is associated with singularity $\xi = \xi_s$ by $\Lambda_s$. For the singular late-order behaviour to be consistent with the local behaviour of the leading-order solution in the neighborhood of the singularity~\eqref{1:u0loc}, we calculate that $\beta_2 = 1$. In Appendix \ref{app:localw}, performing a local expansion of the solutions $u(\xi)$ and $v(\xi)$ in the neighborhood of the singularity, and using asymptotic matching, we determine that
\begin{equation}\label{1:Lambda}
	\Lambda_s = -\frac{\i\bar{A}({\xi_s})  \sqrt{k}}{ c_A}\,,
\end{equation}
where $\bar{A}({\xi_s})$ is given in~\eqref{e:barA}.
We thereby fully determine the asymptotic behaviour of $u_j$ and $v_j$ in the $j \rightarrow \infty$ limit.

%%%%%%%%%%%%%%%%%%%%%%%%%%%%%%%%%%%%%%%%%%%%%%%%%%%%%%%%%
\subsection{Stokes switching}\label{s:woodstokes}

The next step is to truncate the asymptotic series~\eqref{1:asympseries} optimally and to determine the behaviour of the truncation remainder. Truncating the asymptotic series after $N$ terms, we obtain
\begin{equation}\label{3:series}
	u(\xi) = \sum_{j=0}^{N-1} \eta^{2j} u_j(\xi) + S_N(\xi)\,,  \quad  v(\xi) = \sum_{j=0}^{N-1} \eta^{2j} v_j(\xi) + R_N(\xi)\,,
\end{equation}
where $S_N$ and $R_N$ are the remainder terms obtained by truncating the series, and are exponentially small if we optimally truncate the series. We denote the optimal truncation point as $N = N_{\mathrm{opt}}$. The heuristic in~\cite{Boyd1999} shows that $N_{\mathrm{opt}}$ is determined by finding the point where consecutive terms in the series are equal in size. This heuristic gives 
\begin{equation}
N_{\mathrm{opt}}=|\chi|/2\eta+\omega,
\label{e:Nopt}
\end{equation} 
where $\omega\in[0,1)$ is chosen in a way to ensure that $N_{\mathrm{opt}}$ is an integer. Note that $N_{\mathrm{opt}}\to\infty$ as $\eta\to0$.

Inserting~\eqref{3:series} into~\eqref{1:gov1}--\eqref{1:gov2} gives governing equations for $R_N$ and $S_N$.
Using asymptotic matching, it is possible to examine the problem in the neighbourhood of the Stokes curves, which satisfy $\mathrm{Im}(\chi) = 0$ and $\mathrm{Re}(\chi) > 0$. The details of this matching process originate in~\cite{Daalhuis}, and the implementation is very similar to that described in~\cite{Chapman}. We have included the full details in Appendix \ref{a:switching}.

From this analysis, we find that the the total exponentially small contribution switched on across the Stokes curve caused by the singularity pair $\xi_s$ and $\xi_s^*$ is
\begin{equation}
	[R_N]_-^+ \sim\frac{2|\bar{A}(\xi_s)|   \pi\sqrt{k}}{\eta c_A}
\exp\left(-\frac{\sqrt{k}\mathrm{Im}(\xi_s)}{c_A\eta}\right)
\cos\left(\frac{\sqrt{k}(\xi-\mathrm{Re}(\xi_s))}{c_A\eta}-\theta_{\xi_s}\right)\quad \mathrm{as} \quad \eta \rightarrow 0\,,
\label{e:woodpile_sn}
\end{equation}
where $[R_N]_-^+$ denotes the change in the remainder contribution as the Stokes curve is crossed from $\mathrm{Im}(\chi)<0$ to $\mathrm{Im}(\chi)>0$, and $\bar{A}(\xi_s)=|\bar{A}(\xi_s)|\e^{\mathrm{i}\theta_{\xi_s}}$ is defined in~\eqref{e:barA}, and is determined by the coefficients $C_{2n+1}$, the location of singularity $\xi_s$ and the amplitude $A$ of the leading-order wave.
\begin{figure}[tb]
\centering
\includegraphics{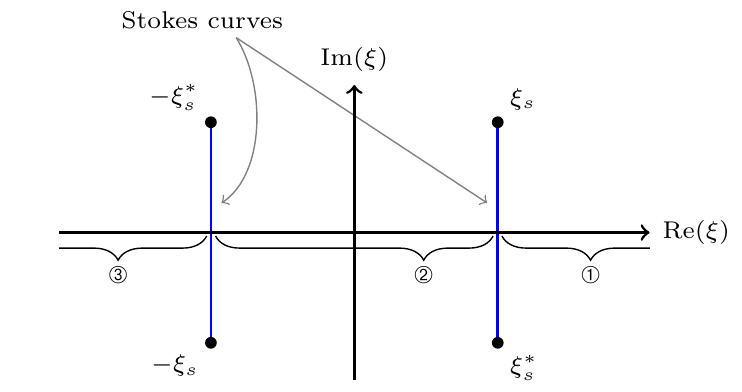}
\caption{Stokes structure associated with the leading-order solitary wave~\eqref{e:senmanciu}. The end points of the Stokes curves are located at singularities of the leading-order solution~\eqref{e:senmanciu}, represented by filled black circles. The Stokes curves are represented by vertical blue lines. The real axis is divided into three regions, denoted \ding{192}, \ding{193}, and \ding{194}. In \ding{192}, no Stokes contributions are present; in \ding{193}, one of the Stokes contributions has been switched on; in \ding{194}, two different Stokes contributions are active, associated with the two Stokes curves. }
\label{f:stokes}
\end{figure}

%%%%%%%%%%%%%%%%%%%%%%%%%%%%%%%%%%%%%%%%%%%%%%%%%%%%%%%%
\subsection{Comparison of our asymptotic and computational results}\label{s:woodnum}

In this section, we determine the oscillatory behaviour following the leading-order wave for specific choices of $\alpha$, and compare our results to numerical experiments.

For each value of $\alpha$, there are two important singularity pairs. Each pair generates a Stokes curve which connects singularities in the pair, intersecting the real axis. As illustrated in Figure~\ref{f:stokes}, the Stokes curves divides the real axis into three regions, denoted \ding{192}, \ding{193}, and \ding{194}.
In \ding{192}, which is ahead of the leading-order solitary wave, there are no exponentially small contributions, and hence no oscillations. Moving into \ding{193} crosses one of the Stokes curves, causing one exponentially small oscillatory contribution to appear. Continuing into \ding{194} crosses a second Stokes curve, causing a second set of exponentially small oscillations to appear in the solution.

Adding the contributions from the two Stokes curves and using trigonometric identities to simplify, we obtain overall amplitude of the exponentially small oscillations when both contributions are present
\begin{align}
\label{e:amp}
\mathrm{Amplitude} \sim \frac{4|\bar{A}({\xi_{s}})|   \pi\sqrt{k}}{\eta c_A}
\exp\left(-\frac{\sqrt{k}\mathrm{Im}(\xi_s)}{c_A\eta}\right)
\cos\left(\frac{\sqrt{k}\mathrm{Re}(\xi_s)}{c_A\eta}+\theta_{\xi_s}\right).
\end{align}

In Table~\ref{t:2}, the values of $\xi_s$, $|\bar{A}({\xi_{s}})|$ and $\theta_{\xi_s}$ are reported for a range of choices of $\alpha$. Note that $|\bar{A}({\xi_{s}})|$ is proportional to $A$, which is the amplitude of the leading-order solitary wave.

\begin{table}[t!]
\caption{Singularities and Values of $|\bar{A}({\xi_{s}})|$ and $\theta_{\xi_s}$ for different values of $\alpha$.}
\label{t:2}
\centering
\begin{tabular}{|c| c c c c| } \hline
 $\alpha$ & $\mathrm{Im}(\xi_{s})$  &  $\mathrm{Re}(\xi_{s})$ & $|\bar{A}({\xi_{s}})|$   &  $\theta_{\xi_s}$ \\ \hline
1.2       & 2.8855  & $0.5598 $ &1.2286A  & -1.5039 \\ \hline
1.5       & 1.8674  & $0.4857$ & 0.9216A  & -1.5470  \\ \hline
1.75      & 1.6503& $0.4728$ &0.8638A  & -1.5031  \\ \hline
2         & 1.4627  & $0.4790$ & 0.6643A  & -1.6290\\ \hline
2.5       & 1.1891  & $0.4947$ & 0.4161A  & -1.5525\\ \hline
3         & 1.0106  & $0.4960$ & 0.2862A  & -1.4757 \\ \hline
3.5       & 0.8967  & $0.4993$ & 0.2328A  & -1.4535 \\ \hline
\end{tabular}
\end{table}

For numerical comparisons, we employ a symplectic integrator using the velocity Verlet algorithm~\cite{Verlet,Allen}.
As in~\cite{Lustri1,Deng2021}, we perform numerical simulations on the relative displacements
\begin{equation}
	r_1(n,t) = u(n+1,t) - u(n,t)\,, \quad  r_2(n,t) = v(n+1,t) - v(n,t)\,.
\end{equation}
In these variables, the leading-order solitary wave decays to zero away from the center in both directions, allowing us to simulate an infinite time domain by truncating the spatial domain and imposing a periodic boundary condition. We performed computations on a woodpile chain with $M=2^{10}$ heavy particles and $M=2^{10}$ light particles in the truncated domain, both with indices $n \in \{ -M/2 + 1, \ldots, M/2\}$. At $t=0$ the woodpile chain is excited by a solitary wave solution of the monatomic Hertzian chain~\eqref{1:woodpile12}. In terms of the relative displacements, the initial state is given as $r_1(n,0)=u_0(n+1,0)-u_0(n,0)$ and $r_2(n,0)=v_0(n+1,0)-v_0(n,0)$ with $u_0$ and $v_0$ given by~\eqref{1:leading} and~\eqref{e:senmanciu}.

To avoid interactions between the tailing oscillations and the leading-order solitary wave we apply the windowing procedure used in~\cite{Giardetti,Lustri1,Deng2021}. At each time step we  multiply $r_1(n,t)$, $\dot{r}_1(n,t)$, $r_2(n,t)$ and $\dot{r}_2(n,t)$ by a function $W(n-n_{\max}+M/8)$, where
\begin{equation}\label{e:window}
W(k) = \left\{
        \begin{array}{ll}
            1\,, & \quad |K| \leq \tfrac{5M}{16} \\
            1 - \tfrac{8}{M}\left(|K| - \tfrac{5M}{16}\right)\,, & \quad \tfrac{5M}{16} < |K| \leq \tfrac{7M}{16} \\
            0\,, & \quad\tfrac{7M}{16} < |K| \leq \tfrac{M}{2}
        \end{array}
    \right.
\end{equation}
and $n_{\max}$ denotes the center of the leading-order solution.

\begin{figure}[tb]
\centering

\subfloat[$\alpha = 1.5$]{
\includegraphics{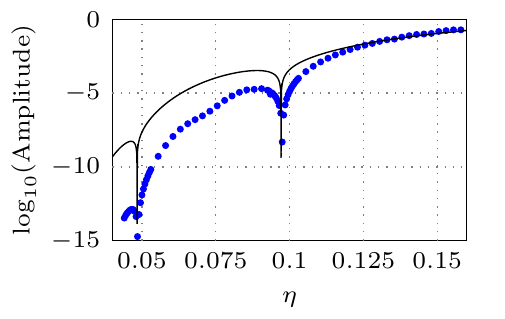}
%\begin{tikzpicture}
%[x =30cm,y =0.15cm]
%\node at (0.16,-10) [white,right] {\scriptsize{$1$}}; % Spacing
%
%\draw (0.04,-15) -- (0.16,-15) -- (0.16,0) -- (0.04,0) -- cycle;
%
%\draw[gray,dotted] (0.04,-5) -- (0.16,-5);
%\draw[gray,dotted] (0.04,-10) -- (0.16,-10);
%\node at (0.04,0) [left] {\scriptsize{$0$}};
%\node at (0.04,-5) [left] {\scriptsize{$-5$}};
%\node at (0.04,-10) [left] {\scriptsize{$-10$}};
%\node at (0.04,-15) [left] {\scriptsize{$-15$}};
%
%\draw [gray,dotted] (0.05,-15) -- (0.05,0);
%\draw [gray,dotted] (0.075,-15) -- (0.075,0);
%\draw [gray,dotted] (0.1,-15) -- (0.1,0);
%\draw [gray,dotted] (0.125,-15) -- (0.125,0);
%\draw [gray,dotted] (0.15,-15) -- (0.15,0);
%\node at (0.05,-15) [below] {\scriptsize{$0.05$}};
%\node at (0.075,-15) [below] {\scriptsize{$0.075$}};
%\node at (0.1,-15) [below] {\scriptsize{$0.1$}};
%\node at (0.125,-15) [below] {\scriptsize{$0.125$}};
%\node at (0.15,-15) [below] {\scriptsize{$0.15$}};
%
%
%\draw[blue] plot[only marks, mark=*, mark size=0.8pt] file {n25khalf_a5.txt};
%\draw[black] plot[smooth,line width=0.75mm] file {n25khalf_a5_asymp.txt};
%
%\draw node at (0.01,-7.5) [rotate=90] {\scriptsize{$\log_{10}(\mathrm{Amplitude})$}};
%\draw node at (0.1,-19) {\scriptsize{$\eta$}};
%
%
%\end{tikzpicture}
}
\subfloat[$\alpha = 1.75$]{
\includegraphics{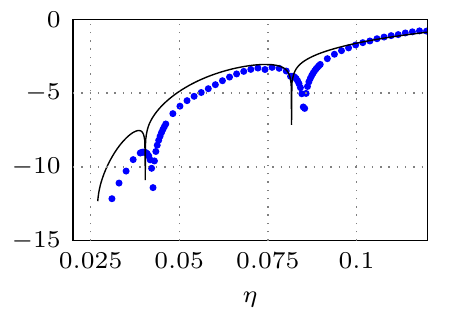}
%\begin{tikzpicture}
%[x =36cm,y =0.15cm]
%
%\node at (0.12,-10) [white,right] {\scriptsize{$1$}}; % Spacing
%
%\draw[blue] plot[only marks, mark=*, mark size=0.8pt] file {n275khalf_a5.txt};
%\draw[black] plot[smooth,line width=0.75mm] file {n275khalf_a5_asymp.txt};
%\fill[white] (0.12,-15) -- (0.125,-15) -- (0.125,0) -- (0.12,0) -- cycle;
%
%\draw (0.02,-15) -- (0.12,-15) -- (0.12,0) -- (0.02,0) -- cycle;
%
%\draw[gray,dotted] (0.02,-5) -- (0.12,-5);
%\draw[gray,dotted] (0.02,-10) -- (0.12,-10);
%\node at (0.02,0) [left] {\scriptsize{$0$}};
%\node at (0.02,-5) [left] {\scriptsize{$-5$}};
%\node at (0.02,-10) [left] {\scriptsize{$-10$}};
%\node at (0.02,-15) [left] {\scriptsize{$-15$}};
%
%\draw [gray,dotted] (0.025,-15) -- (0.025,0);
%\draw [gray,dotted] (0.05,-15) -- (0.05,0);
%\draw [gray,dotted] (0.075,-15) -- (0.075,0);
%\draw [gray,dotted] (0.1,-15) -- (0.1,0);
%\node at (0.025,-15) [below] {\scriptsize{$0.025$}};
%\node at (0.05,-15) [below] {\scriptsize{$0.05$}};
%\node at (0.075,-15) [below] {\scriptsize{$0.075$}};
%\node at (0.1,-15) [below] {\scriptsize{$0.1$}};
%
%
%%\draw node at (-0.01,-7) [rotate=90] {\scriptsize{$\log_{10}(\mathrm{Amplitude})$}};
%\draw node at (0.07,-19) {\scriptsize{$\eta$}};
%
%
%\end{tikzpicture}
}
\subfloat[$\alpha = 2$]{
\includegraphics{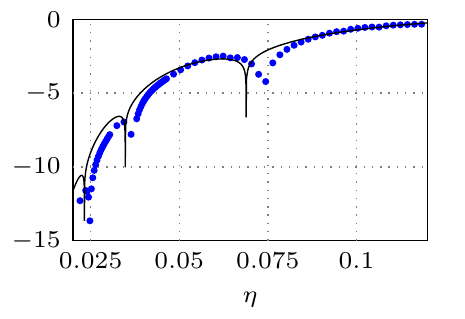}
%\begin{tikzpicture}
%[x =36cm,y =0.15cm]
%\node at (0.12,-10) [white,right] {\scriptsize{$1$}}; % Spacing
%
%\draw (0.02,-15) -- (0.12,-15) -- (0.12,0) -- (0.02,0) -- cycle;
%
%\draw[gray,dotted] (0.02,-5) -- (0.12,-5);
%\draw[gray,dotted] (0.02,-10) -- (0.12,-10);
%\node at (0.02,0) [left] {\scriptsize{$0$}};
%\node at (0.02,-5) [left] {\scriptsize{$-5$}};
%\node at (0.02,-10) [left] {\scriptsize{$-10$}};
%\node at (0.02,-15) [left] {\scriptsize{$-15$}};
%
%\draw [gray,dotted] (0.025,-15) -- (0.025,0);
%\draw [gray,dotted] (0.05,-15) -- (0.05,0);
%\draw [gray,dotted] (0.075,-15) -- (0.075,0);
%\draw [gray,dotted] (0.1,-15) -- (0.1,0);
%\node at (0.025,-15) [below] {\scriptsize{$0.025$}};
%\node at (0.05,-15) [below] {\scriptsize{$0.05$}};
%\node at (0.075,-15) [below] {\scriptsize{$0.075$}};
%\node at (0.1,-15) [below] {\scriptsize{$0.1$}};
%
%
%\draw[blue] plot[only marks, mark=*, mark size=0.8pt] file {n3khalf_a5.txt};
%\draw[black] plot[smooth,line width=0.75mm] file {n3khalf_a5_asymp.txt};
%
%
%%\draw node at (-0.01,-7) [rotate=90] {\scriptsize{$\log_{10}(\mathrm{Amplitude})$}};
%\draw node at (0.07,-19) {\scriptsize{$\eta$}};
%
%
%\end{tikzpicture}
}

\subfloat[$\alpha = 2.5$]{
\includegraphics{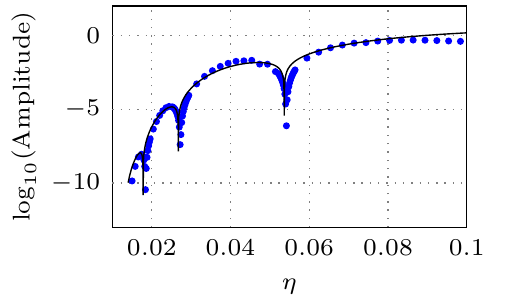}
%\begin{tikzpicture}
%[x =40cm,y =0.15cm]
%\node at (0.1,-10) [white,right] {\scriptsize{$1$}}; % Spacing
%
%\draw (0.01,-13) -- (0.1,-13) -- (0.1,2) -- (0.01,2) -- cycle;
%
%\draw[gray,dotted] (0.01,0) -- (0.1,0);
%\draw[gray,dotted] (0.01,-5) -- (0.1,-5);
%\draw[gray,dotted] (0.01,-10) -- (0.1,-10);
%\node at (0.01,0) [left] {\scriptsize{$0$}};
%\node at (0.01,-5) [left] {\scriptsize{$-5$}};
%\node at (0.01,-10) [left] {\scriptsize{$-10$}};
%
%\draw [gray,dotted] (0.02,-13) -- (0.02,2);
%\draw [gray,dotted] (0.04,-13) -- (0.04,2);
%\draw [gray,dotted] (0.06,-13) -- (0.06,2);
%\draw [gray,dotted] (0.08,-13) -- (0.08,2);
%\node at (0.02,-13) [below] {\scriptsize{$0.02$}};
%\node at (0.04,-13) [below] {\scriptsize{$0.04$}};
%\node at (0.06,-13) [below] {\scriptsize{$0.06$}};
%\node at (0.08,-13) [below] {\scriptsize{$0.08$}};
%\node at (0.1,-13) [below] {\scriptsize{$0.1$}};
%
%\draw[blue] plot[only marks, mark=*, mark size=0.8pt] file {n35khalf_a5.txt};
%\draw[black] plot[smooth,line width=0.75mm] file {n35khalf_a5_asymp.txt};
%
%
%\draw node at (-0.0125,-5.5) [rotate=90] {\scriptsize{$\log_{10}(\mathrm{Amplitude})$}};
%\draw node at (0.055,-17) {\scriptsize{$\eta$}};
%
%
%\end{tikzpicture}
}
\subfloat[$\alpha = 3$]{
\includegraphics{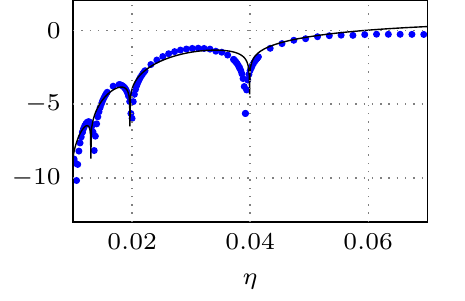}
%\begin{tikzpicture}
%[x =60cm,y =0.15cm]
%\node at (0.07,-10) [white,right] {\scriptsize{$1$}}; % Spacing
%
%\draw[blue] plot[only marks, mark=*, mark size=0.8pt] file {n4khalf_a5.txt};
%\draw[black] plot[smooth,line width=0.75mm] file {n4khalf_a5_asymp.txt};
%
%\fill[white] (0.01,-12) -- (0.005,-12) -- (0.005,2) -- (0.01,2) -- cycle;
%
%\draw (0.01,-13) -- (0.07,-13) -- (0.07,2) -- (0.01,2) -- cycle;
%
%\draw[gray,dotted] (0.01,0) -- (0.07,0);
%\draw[gray,dotted] (0.01,-5) -- (0.07,-5);
%\draw[gray,dotted] (0.01,-10) -- (0.07,-10);
%\node at (0.01,0) [left] {\scriptsize{$0$}};
%\node at (0.01,-5) [left] {\scriptsize{$-5$}};
%\node at (0.01,-10) [left] {\scriptsize{$-10$}};
%
%\draw [gray,dotted] (0.02,-13) -- (0.02,2);
%\draw [gray,dotted] (0.04,-13) -- (0.04,2);
%\draw [gray,dotted] (0.06,-13) -- (0.06,2);
%\node at (0.02,-13) [below] {\scriptsize{$0.02$}};
%\node at (0.04,-13) [below] {\scriptsize{$0.04$}};
%\node at (0.06,-13) [below] {\scriptsize{$0.06$}};
%
%%draw node at (-0.005,-7) [rotate=90] {\scriptsize{$\log_{10}(\mathrm{Amplitude})$}};
%\draw node at (0.04,-17) {\scriptsize{$\eta$}};
%
%\end{tikzpicture}
}
\subfloat[$\alpha = 3.5$]{
\includegraphics{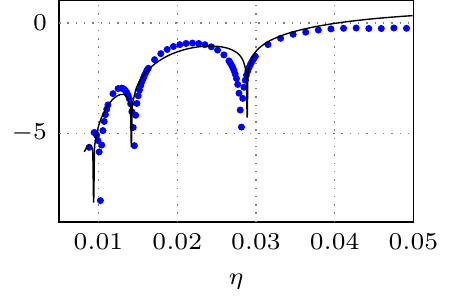}
%\begin{tikzpicture}
%[x =80cm,y =0.225cm]
%\node at (0.05,-5) [white,right] {\scriptsize{$1$}}; % Spacing
%
%\draw (0.005,-9) -- (0.05,-9) -- (0.05,1) -- (0.005,1) -- cycle;
%
%\draw[gray,dotted] (0.005,0) -- (0.05,0);
%\draw[gray,dotted] (0.005,-5) -- (0.05,-5);
%\node at (0.005,0) [left] {\scriptsize{$0$}};
%\node at (0.005,-5) [left] {\scriptsize{$-5$}};
%
%\draw [gray,dotted] (0.02,-9) -- (0.02,1);
%\draw [gray,dotted] (0.04,-9) -- (0.04,1);
%\draw [gray,dotted] (0.03,-9) -- (0.03,1);
%\draw [gray,dotted] (0.01,-9) -- (0.01,1);
%\node at (0.02,-9) [below] {\scriptsize{$0.02$}};
%\node at (0.04,-9) [below] {\scriptsize{$0.04$}};
%\node at (0.03,-9) [below] {\scriptsize{$0.03$}};
%\node at (0.01,-9) [below] {\scriptsize{$0.01$}};
%\node at (0.05,-9) [below] {\scriptsize{$0.05$}};
%
%\draw[blue] plot[only marks, mark=*, mark size=0.8pt] file {n45khalf_a5.txt};
%\draw[black] plot[smooth,line width=0.75mm] file {n45khalf_a5_asymp.txt};
%
%
%%\draw node at (-0.005,-5) [rotate=90] {\scriptsize{$\log_{10}(\mathrm{Amplitude})$}};
%\draw node at (0.0275,-11.66667) {\scriptsize{$\eta$}};
%
%\end{tikzpicture}
}
\caption{Comparison between the asymptotic results  of the tailing oscillations~\eqref{e:amp} and the numerical results  for values of $\alpha=1.5,1.75,2,2.5,3,3.5$.  The amplitude of the leading wave is fixed to be 5 and the spring constant $k$ is fixed to be $0.5$. The comparison indicates that the asymptotic approximation is an effective method for predicting both the oscillation amplitude and the anti-resonance points, especially for larger values of $\alpha$.}\label{f:compare1}
\end{figure}
Using~\eqref{e:amp}, we obtain the amplitude of the total oscillations for each $\alpha$. These values are compared with numerical simulations in Figure~\ref{f:compare1}. In each of these figures, there are anti-resonance conditions, or values of the mass ratio for which the oscillating tail vanishes. These anti-resonance values were not present in the weakly nonlinear regime, and are therefore caused by nonlinear effects in the system. The asymptotic prediction of these anti-resonant mass ratios is consistent with the numerically obtained values for all values of $\alpha$.

Our prediction of the amplitude of the oscillating tail improves as $\alpha$ increases, with significant errors present for $\alpha = 1.5$. As indicated in~\eqref{e:amp} the amplitude depends exponentially on the imaginary part of the singularity locations, which are determined by the values of $C_{2n+1}$ reported in Table~\ref{t:1}. Note that for smaller $\alpha$, $C_5$ is particularly small; this is notable in the case $\alpha = 1.5$.

Recall that $C_5$ is determined by taking the fifth time derivative of the displacement of one particle, and taking repeated numerical differentiations has the effect of magnifying accumulated numerical inaccuracy. This is significant when the computed value is small. 

\begin{figure}[tb]
\centering

\subfloat[$\alpha = 1.2$]{
\includegraphics{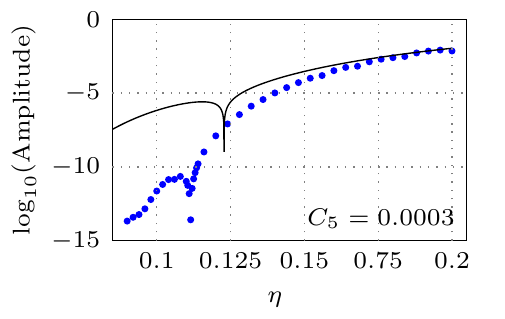}
%\begin{tikzpicture}
%[x =30cm,y =0.15cm]
%\node at (0.205,-8) [white,right] {\scriptsize{$1$}}; % Spacing
%\draw[blue] plot[only marks, mark=*, mark size=0.8pt] file {n22khalf_a5.txt};
%\draw[black] plot[smooth,line width=0.75mm] file {n22khalf_a5_asymp.txt};
%\fill[white] (0.085,-15) -- (0.079,-15) -- (0.079,0) -- (0.085,0) -- cycle;
%
%\draw (0.085,-15) -- (0.205,-15) -- (0.205,0) -- (0.085,0) -- cycle;
%
%\draw[gray,dotted] (0.085,-5) -- (0.205,-5);
%\draw[gray,dotted] (0.085,-10) -- (0.205,-10);
%\node at (0.085,-5) [left] {\scriptsize{$-5$}};
%\node at (0.085,-10) [left] {\scriptsize{$-10$}};
%\node at (0.085,0) [left] {\scriptsize{$0$}};
%\node at (0.085,-15) [left] {\scriptsize{$-15$}};
%
%\draw [gray,dotted] (0.1,-15) -- (0.1,0);
%\draw [gray,dotted] (0.125,-15) -- (0.125,0);
%\draw [gray,dotted] (0.15,-15) -- (0.15,0);
%\draw [gray,dotted] (0.175,-15) -- (0.175,0);
%\draw [gray,dotted] (0.2,-15) -- (0.2,0);
%\node at (0.1,-15) [below] {\scriptsize{$0.1$}};
%\node at (0.125,-15) [below] {\scriptsize{$0.125$}};
%\node at (0.15,-15) [below] {\scriptsize{$0.15$}};
%\node at (0.175,-15) [below] {\scriptsize{$0.75$}};
%\node at (0.2,-15) [below] {\scriptsize{$0.2$}};
%
%\node at(0.205,-15) [above left] {\scriptsize{ $C_5=3.26\times10^{-4}$}};
%
%
%
%
%\draw node at (0.055,-7.5) [rotate=90] {\scriptsize{$\log_{10}(\mathrm{Amplitude})$}};
%\draw node at (0.14,-19) {\scriptsize{$\eta$}};
%
%
%\end{tikzpicture}
}
\subfloat[$\alpha = 1.5$]{
\includegraphics{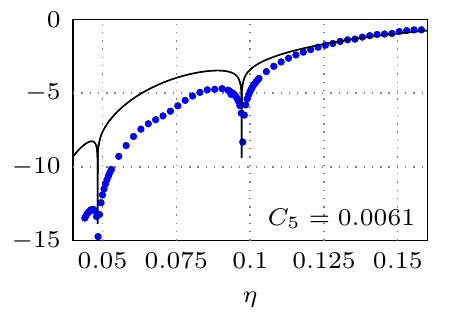}
%\begin{tikzpicture}
%[x =30cm,y =0.15cm]
%\node at (0.16,-10) [white,right] {\scriptsize{$1$}}; % Spacing
%
%\draw (0.04,-15) -- (0.16,-15) -- (0.16,0) -- (0.04,0) -- cycle;
%
%\draw[gray,dotted] (0.04,-5) -- (0.16,-5);
%\draw[gray,dotted] (0.04,-10) -- (0.16,-10);
%\node at (0.04,0) [left] {\scriptsize{$0$}};
%\node at (0.04,-5) [left] {\scriptsize{$-5$}};
%\node at (0.04,-10) [left] {\scriptsize{$-10$}};
%\node at (0.04,-15) [left] {\scriptsize{$-15$}};
%
%\draw [gray,dotted] (0.05,-15) -- (0.05,0);
%\draw [gray,dotted] (0.075,-15) -- (0.075,0);
%\draw [gray,dotted] (0.1,-15) -- (0.1,0);
%\draw [gray,dotted] (0.125,-15) -- (0.125,0);
%\draw [gray,dotted] (0.15,-15) -- (0.15,0);
%\node at (0.05,-15) [below] {\scriptsize{$0.05$}};
%\node at (0.075,-15) [below] {\scriptsize{$0.075$}};
%\node at (0.1,-15) [below] {\scriptsize{$0.1$}};
%\node at (0.125,-15) [below] {\scriptsize{$0.125$}};
%\node at (0.15,-15) [below] {\scriptsize{$0.15$}};
%
%
%\draw[blue] plot[only marks, mark=*, mark size=0.8pt] file {n25khalf_a5.txt};
%\draw[black] plot[smooth,line width=0.75mm] file {n25khalf_a5_asymp.txt};
%
%%\draw node at (0.01,-7.5) [rotate=90] {\scriptsize{$\log_{10}(\mathrm{Amplitude})$}};
%\draw node at (0.1,-19) {\scriptsize{$\eta$}};
%
%\node at(0.16,-15) [above left] {\scriptsize{ $C_5=6.13\times10^{-3}$}};
%
%\draw[blue] plot[only marks, mark=*, mark size=0.8pt] file {n25khalf_a5.txt};
%\draw[black] plot[smooth,line width=0.75mm] file {n25khalf_a5_asymp.txt};
%
%
%
%\end{tikzpicture}
}
\subfloat[$\alpha = 2$ ]{
\includegraphics{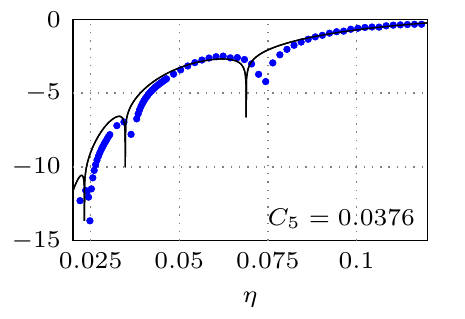}
%\begin{tikzpicture}
%[x =36cm,y =0.15cm]
%\node at (0.12,-10) [white,right] {\scriptsize{$1$}}; % Spacing
%
%\draw (0.02,-15) -- (0.12,-15) -- (0.12,0) -- (0.02,0) -- cycle;
%
%\draw[gray,dotted] (0.02,-5) -- (0.12,-5);
%\draw[gray,dotted] (0.02,-10) -- (0.12,-10);
%\node at (0.02,0) [left] {\scriptsize{$0$}};
%\node at (0.02,-5) [left] {\scriptsize{$-5$}};
%\node at (0.02,-10) [left] {\scriptsize{$-10$}};
%\node at (0.02,-15) [left] {\scriptsize{$-15$}};
%
%\draw [gray,dotted] (0.025,-15) -- (0.025,0);
%\draw [gray,dotted] (0.05,-15) -- (0.05,0);
%\draw [gray,dotted] (0.075,-15) -- (0.075,0);
%\draw [gray,dotted] (0.1,-15) -- (0.1,0);
%\node at (0.025,-15) [below] {\scriptsize{$0.025$}};
%\node at (0.05,-15) [below] {\scriptsize{$0.05$}};
%\node at (0.075,-15) [below] {\scriptsize{$0.075$}};
%\node at (0.1,-15) [below] {\scriptsize{$0.1$}};
%
%
%\draw[blue] plot[only marks, mark=*, mark size=0.8pt] file {n3khalf_a5.txt};
%\draw[black] plot[smooth,line width=0.75mm] file {n3khalf_a5_asymp.txt};
%
%
%%\draw node at (-0.01,-7) [rotate=90] {\scriptsize{$\log_{10}(\mathrm{Amplitude})$}};
%\draw node at (0.07,-19) {\scriptsize{$\eta$}};
%
%\node at(0.12,-15) [above left] {\scriptsize{ $C_5=3.76\times10^{-2}$}};
%
%\draw[blue] plot[only marks, mark=*, mark size=0.8pt] file {n3khalf_a5.txt};
%\draw[black] plot[smooth,line width=0.75mm] file {n3khalf_a5_asymp.txt};
%
%
%
%\end{tikzpicture}
}\caption{Comparison between asymptotic approximation and numerics for values of $\alpha=1.2,1.5,2$, where the coefficient $C_5$ in~\eqref{e:senmanciu} is relatively small. This comparison supports the conjecture that smaller values of $C_5$ correspond to more significant errors when this approximation method is applied. It is notable that the anti-resonance conditions are still accurately predicted for $\alpha = 1.5$.}
\label{f:comparebad}
\end{figure}

We conjecture that the small value of $C_5$ is responsible for the errors seen at $\alpha = 1.5$. We support this conjecture by including the numerical comparison for $\alpha = 1.2$. In this case, $C_5 \approx 3\times 10^{-4}$~\cite{sen2001}, which is significantly smaller even than the corresponding value for $\alpha = 1.5$. We see in Figure~\ref{f:comparebad} that the asymptotic prediction of the oscillation amplitude for $\alpha = 1.2$ is significantly less accurate than that for $\alpha = 1.5$.

We suggest that these errors could be averted by applying a different approximation for the leading-order behaviour, such as rational approximations~\cite{pade,Nakatsukasa}, which can approximate the leading-order behaviour to arbitrary precision, although this is beyond the scope of the present study. Even despite these errors in the amplitude calculation for smaller values of $\alpha$, the comparisons in Figure~\ref{f:compare1} show that this method is particularly effective at identifying anti-resonance conditions, and can accurately predict the oscillation amplitudes for a wide range of system parameters.

The asymptotic prediction of the anti-resonance condition, corresponding to the cancellation of the tailing oscillations, can be derived from~\eqref{e:amp}
\begin{equation}
\frac{\sqrt{k}\mathrm{Re}(\xi_s)}{c_A\eta}=\bigg(n+\frac12\bigg)\pi-\theta_{\xi_s}.
\label{e:resonance1}
\end{equation}

The anti-resonance condition was identified for the zero precompression woodpile chain in~\cite{Xu} by considering the Fourier transform of system~\eqref{1:gov1}--\eqref{1:gov2} over the whole real line. They demonstrated this transform is well defined only for a set of system parameters. The existence of the Fourier transform implies that the solution decays to zero at infinity, corresponding to the anti-resonance condition. This condition is given in~\cite{Xu} as
\begin{equation}
\frac{\sqrt{k(1+\eta)}}{c_A\eta}=2n\pi.
\label{e:resonance2}
\end{equation}
 We note that the numerically-derived quantities $\mathrm{Re}(\xi_s) \approx 0.5$ and $\theta_{\xi_s} \approx \pi/2$. Hence, in the limit $\eta\to0$,~\eqref{e:resonance1} is consistent with~\eqref{e:resonance2}.

%%%%%%%%%%%%%%%%%%%%%%%%%%%%%%%%%%%%%%%%%%%%%%%%%%%%%%%%%%%%%%%%%%%%%%%%%%%%%%%%%%%%%%%%%%%%%%%%%
%%%%%%%%%%%%%%%%%%%%%%%%%%%%%%%%%%%%%%%%%%%%%%%%%%%%%%%%%%%%%%%%%%%%%%%%%%%%%%%%%%%%%%%%%%%%%%%%%
\section{Transition from the strongly nonlinear regime to weakly nonlinear regime}\label{s:transition}
Section~\ref{s:woodnum} showed that the zero precompression regime allows for chain configurations which produce anti-resonance conditions, and hence localized solitary waves. These configurations do not exist for the strongly precompressed regime~\cite{Deng2021}.

\begin{figure}[b!]
\centering
\includegraphics{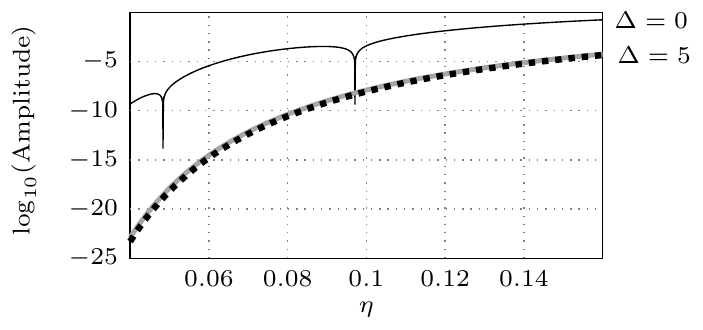}
%{
%\begin{tikzpicture}
%[x =40cm,y =0.1cm]
%\draw (0.04,-25) -- (0.16,-25) -- (0.16,0) -- (0.04,0) -- cycle;
%
%\draw[gray,dotted] (0.04,-5) -- (0.16,-5);
%\draw[gray,dotted] (0.04,-10) -- (0.16,-10);
%\draw[gray,dotted] (0.04,-15) -- (0.16,-15);
%\draw[gray,dotted] (0.04,-20) -- (0.16,-20);
%\node at (0.04,-5) [left] {\scriptsize{$-5$}};
%\node at (0.04,-10) [left] {\scriptsize{$-10$}};
%\node at (0.04,-15) [left] {\scriptsize{$-15$}};
%\node at (0.04,-20) [left] {\scriptsize{$-20$}};
%\node at (0.04,-25) [left] {\scriptsize{$-25$}};
%
%\draw [gray,dotted] (0.06,-25) -- (0.06,0);
%\draw [gray,dotted] (0.08,-25) -- (0.08,0);
%\draw [gray,dotted] (0.1,-25) -- (0.1,0);
%\draw [gray,dotted] (0.12,-25) -- (0.12,0);
%\draw [gray,dotted] (0.14,-25) -- (0.14,0);
%\node at (0.06,-25) [below] {\scriptsize{$0.06$}};
%\node at (0.08,-25) [below] {\scriptsize{$0.08$}};
%\node at (0.1,-25) [below] {\scriptsize{$0.1$}};
%\node at (0.12,-25) [below] {\scriptsize{$0.12$}};
%\node at (0.14,-25) [below] {\scriptsize{$0.14$}};
%
%
%\draw[black] plot[smooth,line width=0.75mm] file {n25khalf_a5_asymp.txt} node[right] {\scriptsize{$\Delta = 0$}};
%\draw[densely dashed,red] plot[smooth,line width=1mm] file {n25_LargeDelta.txt} node[right,black] {\scriptsize{$\Delta = 5$}};
%\draw[blue] plot[smooth,line width=0.45mm] file {n25_LargeDelta_num.txt};
%
%\draw node at (0.013,-12) [rotate=90] {\scriptsize{$\log_{10}(\mathrm{Amplitude})$}};
%\draw node at (0.1,-30.25) {\scriptsize{$\eta$}};
%
%\end{tikzpicture}
%}

\caption{The amplitude of the tailing oscillations as a function of $\eta$ in the woodpile chain with zero precompression (thin black curve) and large precompression (weakly nonlinear asymptotic calculation in gray, nonlinear calculation with numerically computed leading order in dotted black). The amplitude of the leading-order solitary wave is 5, the spring constant is 0.5 and the precompression in the strongly precompressed case is 5.}
\label{f:comparison}
\end{figure}

In Figure~\ref{f:comparison}, we show the amplitude of oscillations that form behind a solitary wave of amplitude 5 with $\alpha = 1.5$ as a function of $\eta$ for $\Delta=0$ and $\Delta=5$. It is apparent that in the former case, there are values of the mass ratio in which the oscillations cancel, but this does not occur when the chain is under precompression. 

For a woodpile chain with zero precompression, the amplitude of the tailing oscillation is given by~\eqref{e:amp}. For a woodpile chain with large precompression, which is in a weakly nonlinear regime, we transform the woodpile chain into the KdV equation to obtain the leading-order solution
\begin{equation}
u_0(\xi)=-\frac{\Delta\epsilon}{\alpha-1}\tanh(\epsilon\xi),\quad v_0(\xi)=u_0(\xi),
\quad \xi = n-c_{\epsilon}t,
\label{e:leading_transition}
\end{equation}
where $\epsilon$ quantifies the nonlinearity of the woodpile chain and $c_\epsilon=\sqrt{\alpha}\Delta^{(\alpha-1)/2}+\sqrt{\alpha}\Delta^{(\alpha-1)/2}\epsilon^2/6$ is the velocity of the leading-order wave. For $\alpha=1.5$, $\Delta=5$ and $\epsilon= 0.25$ corresponds to the leading-order wave amplitude of 5. For more details on this calculation see~\cite{Nesterenko1}.
Performing an exponential asymptotic analysis on~\eqref{e:leading_transition}, the amplitude of the tailing oscillation in the weakly nonlinear regime is obtained in~\cite{Deng2021} as
\begin{equation}
\mathrm{Amplitude}_{\mathrm{asymp}}\sim\frac{2\Delta\pi\sqrt{k}}{(\alpha-1)\eta c_{\epsilon}}\exp\bigg(-\frac{\pi\sqrt{k}}{2c_{\epsilon}\epsilon\eta}\bigg).
\label{e:tailing1}
\end{equation}

For $\Delta=5$, we can also determine the leading-order solution numerically and obtain the tailing oscillations by performing exponential asymptotic analysis on the numerical leading-order solution.

To generate a numerical leading-order solitary wave, we first apply a velocity impulse to a simulated homogeneous chain. We note that, while it is straightforward to fix the amplitude for a zero compression solitary wave using the scaling relation between the impulse velocity and the solitary wave amplitude~\cite{Nesterenko1,Deng1},  it is more challenging to fix the amplitude if the system contains precompression. In this case, the relation between the velocity impulse and the wave amplitude is not straightforward. 
In our simulations, the magnitude of the velocity impulse was carefully tuned to obtain a solitary wave solution with the desired amplitude.

\begin{figure}[b!]
\centering
\includegraphics{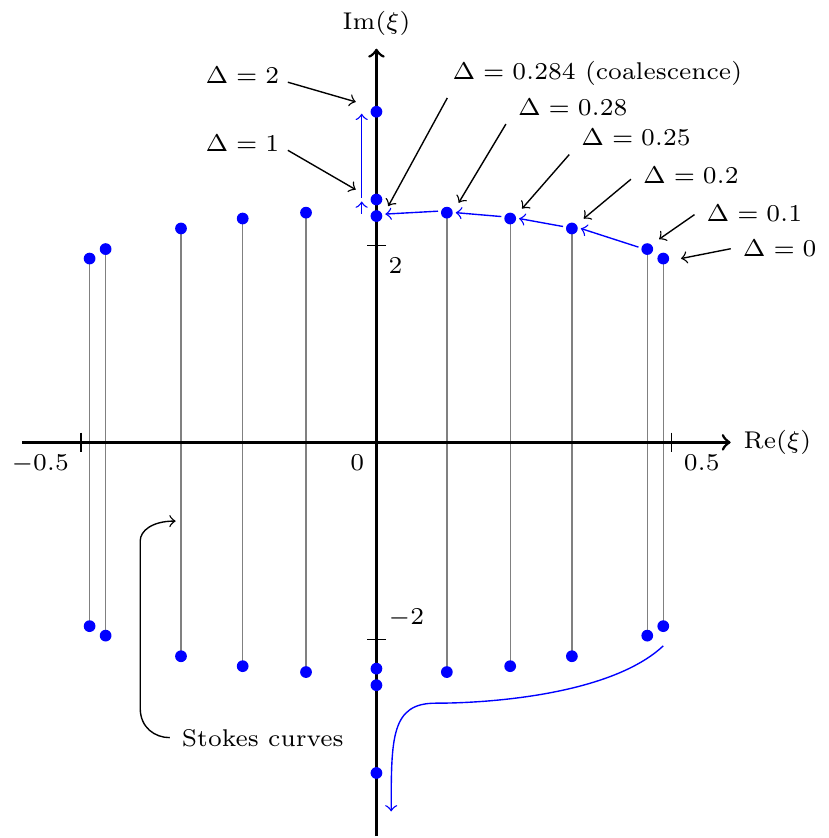}
\caption{The singularity pairs in the complex plane for different values of the precompression $\Delta$, where the leading-order solitary wave has amplitude 5. Stokes lines extend vertically from the singularities, connecting the conjugate pairs. As the precompression increases, the singularities move towards the imaginary axis, eventually coalescing. For small values of the precompression, the solution contains two Stokes curves, each of which produces an oscillatory contribution. Once the value of the precompression exceeds a particular value (in this case, approximately 0.284), the singularity pairs coalesce, and there is only a single Stokes curve presents in the solution, leading to a single oscillatory contribution to the solution. This explains why the oscillations can cancel entirely if the precompression is zero or small, but this does not occur in the strong precompression case considered in \cite{Deng2021}}
\label{f:singularity}
\end{figure}

Assuming that the leading-order solution takes the form of~\eqref{e:senmanciu}, the coefficients $C_{2n+1}$ are obtained following the same fashion described in Section~\ref{s:woodLO}.
Performing the exponential asymptotic analysis on this numerically determined leading-order solution, we calculate the amplitude of the tailing oscillation following the leading-order wave of amplitude 5
\begin{equation}
\mathrm{Amplitude}_{\mathrm{approx}}\sim\frac{19.6540\pi\sqrt{k}}{\eta c}\exp\bigg(-\frac{6.1745\sqrt{k}}{\eta c}\bigg),
\label{e:tailing2}
\end{equation}
where $c$ is the numerically determined velocity of the leading-order wave. We compare the result in~\eqref{e:tailing2} determined based on the numerical leading-order solution with the result in~\eqref{e:tailing1} determined based on the asymptotic leading-order solution in Figure~\ref{f:comparison}, showing that these methods produce consistent results. This confirms that exponential asymptotic methods based on a numerical leading-order travelling wave can produce accurate results in both weakly and strongly nonlinear regimes.

The difference in anti-resonance behaviour between the zero and strong precompression regimes can be understood by studying singularities in the analytic continuation of the leading-order solution. Singularity pairs in the analytically continued solutions are connected by Stokes curves, each of which generates oscillations in the wake of the leading-order travelling wave. In the zero precompression case, there are two singularity pairs, and hence two Stokes curves. The two curves generate two sets of oscillations, which can cancel precisely and cause anti-resonances. In the strongly precompressed case, there is only a single Stokes curve in the travelling wave solution, and hence only a single oscillatory contribution that can never be cancelled.

To understand the transition between these regimes, we therefore study the behaviour of singularities in the analytic continuation of the leading-order travelling wave for woodpile chains, where we fix the amplitude of the leading-order wave and increase the precompression. 

Figure~\ref{f:singularity} shows the leading-order singularity locations for a range of $\Delta$ values, with the amplitude fixed to be $5$. As $\Delta$ increases, the singularity pairs move towards the imaginary axis. The precompression eventually reaches a critical value $\Delta_c\approx 0.284$, at which the singularities coalesce, and the system no longer permits anti-resonance conditions. 

\begin{figure}[t!]
\centering
\includegraphics{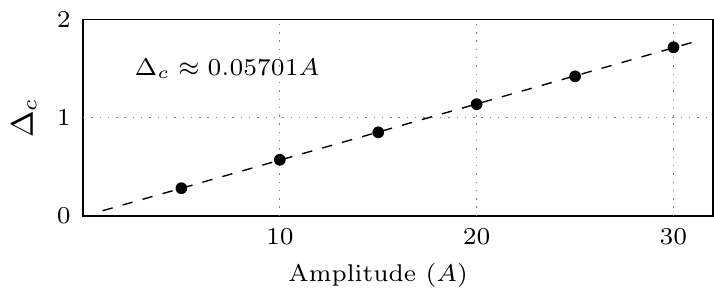}
%{
%\begin{tikzpicture}
%[x =0.2cm,y =1cm]
%\draw (0,0) -- (32,-0) -- (32,2) -- (0,2) -- cycle;
%\draw[gray,dotted] (0,1) -- (32,1);
%\draw[dashed] (1,0.0554) -- (31,1.7680);
%\node at (0,0) [left] {\scriptsize{$0$}};
%\node at (0,1) [left] {\scriptsize{$1$}};
%\node at (0,2) [left] {\scriptsize{$2$}};
%\draw[gray,dotted] (10,0) -- (10,2);
%\node at (10,0) [below] {\scriptsize{$10$}};
%\draw[gray,dotted] (20,0) -- (20,2);
%\node at (20,0) [below] {\scriptsize{$20$}};
%\draw[gray,dotted] (30,0) -- (30,2);
%\node at (30,0) [below] {\scriptsize{$30$}};
%\node at (5, 0.284)[circle,fill,inner sep=1.2pt]{};
%\node at (10, 0.573)[circle,fill,inner sep=1.2pt]{};
%\node at (15, 0.852)[circle,fill,inner sep=1.2pt]{};
%\node at (20, 1.138)[circle,fill,inner sep=1.2pt]{};
%\node at (25, 1.421)[circle,fill,inner sep=1.2pt]{};
%\node at (30, 1.716)[circle,fill,inner sep=1.2pt]{};
%\draw[black] plot[smooth,line width=0.75mm] file {fitting_no_intercept.txt};
%
%\node at (2,1.5)[right]{\scriptsize{$\Delta_c\approx 0.05701A$}};
%
%
%\draw node at (-3,1) [rotate=90] {\small{$\Delta_c$}};
%\draw node at (15,-0.6) {\scriptsize{Amplitude ($A$)}};
%\end{tikzpicture}
%}
\caption{The critical precompression $\Delta_c$ at which the Stokes curves coalesce, as a function of the amplitude of the leading-order wave.}
\label{f:Delta0}
\end{figure}

We then studied the singularity behaviour for leading-order waves with amplitudes 10, 15, 20, 25, and 30. In each case, there are two singularity pairs in the zero precompression leading-order solution. As the precompression increases, the singularity pairs move towards the imaginary axis, and eventually coalesce at some critical value $\Delta_c$ of the precompression. This critical value depends on the leading-order wave amplitude, as shown in Figure~\ref{f:Delta0}. We note that the relationship between $\Delta_c$ and the leading-order amplitude appears to be approximately linear, with the least-squares fitted line shown in Figure~\ref{f:Delta0}

%%%%%%%%%%%%%%%%%%%%%%%%%%%%%%%%%%%%%%%%%%%%%%%%%%%%%%%%%%%%%%%%%%%%%%%%%%%%%%%%%%%%%%%%%%%%%%%%%%
\section{Conclusions and Discussion}
\label{s:conclusion}
In this paper, we investigated %derived asymptotic approximations for 
travelling waves in a woodpile chain, which can be considered as a singularly perturbed granular particle chain. We considered the small mass ratio limit with zero precompression, corresponding to the strongly nonlinear regime. Typical travelling wave solutions in this system are nanoptera, which consist of an exponentially localized central solitary wave and non-decaying exponentially small oscillations in the tail. In this paper, we obtain an explicit asymptotic form of the travelling wave in the zero precompression regime. Existing studies on this system showed that the oscillations in the tail of the leading-order wave vanish for particular mass ratios~\cite{Xu}; however, previous exponential asymptotic analysis on the weakly nonlinear precompressed system did not exhibit this behaviour~\cite{Deng2021}. This paper also connects these two observations by showing that the oscillations do vanish for particular mass ratios in the zero precompression regime, and illustrating how this behaviour relates to that seen in the weakly nonlinear regime.

We found that the analytically-continued leading-order approximation in the zero precompression regime contains two significant Stokes curves, which generate two sets of oscillations with same amplitude but different phases in the tail of the central wave. 
Using exponential asymptotic analysis based on a hybrid numerical-analytic leading-order solution, we obtained an asymptotic form for these oscillations, and we used this form to determine the amplitude of the non-decaying far-field waves, 
which is a superposition of these two sets of oscillations. 
We demonstrated that there exists a set of mass ratios for which two oscillations are precisely out of phase, therefore cancelling entirely. 

In order to demonstrate that these results are consistent with known asymptotic behaviour in the weakly nonlinear regime~\cite{Deng2021}, we studied the behaviour of the leading-order singularities as we increase precompression in the model. For small values of precompression, there are two distinct singularity pairs, generating two Stokes curves in the solution. As the precompression increases and the system approaches the weakly nonlinear regime, the singularity pairs become closer and eventually coalesce. This explains why the leading-order solution in the weakly nonlinear regime only possesses one singularity pair, and hence the solution contains only a single Stokes curve. Because of the coalescing of the Stokes curves, the cancellation of the tailing oscillations is no longer possible in the weakly nonlinear regime, as there is only one set of oscillations generated by the single Stokes curve which can never be counteracted.

In~\cite{Xu} the anti-resonance condition was identified for the woodpile chain with zero precompression by considering the Fourier transform of~\eqref{1:gov1}--\eqref{1:gov2} over the whole real line. This study showed that the Fourier transform over the whole real line is well defined only for a set of system parameters, producing the anti-resonance condition.  The anti-resonance condition determined in~\cite{Xu} agrees with our asymptotic results in the small mass ratio limit. Our method extends on the Fourier existence analysis of~\cite{Xu} by asymptotically calculating the oscillation behaviour, and providing a convenient expression for the amplitude. Our method shows that the tailing oscillations are a consequence of the Stokes phenomenon, providing a mathematical explanation for the appearance of oscillations. We note that exponential asymptotics can be applied to study precompressed chains as seen in 
\cite{Deng2021}, while the Fourier method is not applicable when $\Delta\neq0$. 

A significant challenge to studying the strongly nonlinear regime was the lack of an analytic expression for the leading-order solitary wave. The leading-order solution plays an important role in the exponential asymptotic analysis we use to calculate nanopteron solutions, as singularities in the analytically-continued leading-order solution generate the late-order terms of the asymptotic expansion. In place of an analytic expression for the leading-order behaviour, we used an approximation from in~\cite{sen2001}, which was obtained using a hybrid numerical--analytical method. The success of this approach suggests a path to study a wide range of travelling waves for which the leading-order behaviour can only be obtained numerically, such as chains of magnetic beads which have interactions that extend beyond nearest-neighbour potentials~\cite{Moleron}.

Different approximation methods can potentially give rise to different singularity locations and strengths in the analytic continuation, changing the predicted oscillation behaviour. 
It would be a particularly valuable research direction to study the robustness and accuracy of exponential asymptotic techniques based on different approximation methods. It would be interesting to apply rational approximation~\cite{pade,Nakatsukasa} to obtain the leading-order behaviour, as such methods can be applied with arbitrary precision, and could lead to a highly generic hybrid numerical-analytic technique.

It would be interesting to study other types of singularly perturbed Hertzian chains, such as  diatomic Hertzian chains, in the zero precompression regime. The Hertzian potential~\eqref{e:potential} has a discontinuity that becomes significant when adjacent particles lose contact. In the weakly nonlinear case, each particle stays in contact with its neighbors and the discontinuity of the potential has no effect on the system behaviour. In the zero precompression case, we must consider the effects of this discontinuity. While this not affect the singulant equation~\eqref{e:singulant_woodpile} for woodpile chains, it does impact the singulant expression for diatomic Hertzian chains. We therefore expect that the singulant expression, and subsequent exponential asymptotic analysis, will be significantly more complicated for diatomic chains. 

\section{Acknowledgements} 
This work is under the support of the Australian Research Council Discovery Project DP190101190. The authors thank Professor Mason A. Porter for insightful discussions. 

\appendix
\section{Determining the Prefactor Constants}
\label{app:localw}

We determine the prefactor constants for the woodpile chain. Recall the late-order ansatz~\eqref{2:ansatz},  in a narrow neighborhood of the singularities the earlier terms in the series are not larger asymptotically than later series terms in the limit $\eta \rightarrow 0$.
Therefore the power series ceases to be asymptotic. 
We first obtain a local expansion of the solution near the singular point. Then matching the local solution in the inner region near the singularity with the late-order expansion in the outer region by using Van Dyke's matching principle, we obtain the prefactor constants in the late-order terms.

First we determine the local behaviour of the leading-order solution near singularity $\xi_s$. As $\xi \rightarrow \xi_{s}$, we find that
\begin{align}
	u_0(\xi) &\sim -\frac{\bar{A}(\xi_s)}{\xi - \xi_{s}} + \mathcal{O}(\xi - \xi_{s})\,, \quad & u_0(\xi + 1) &\sim A\tanh(C_1(\xi_s+1)+C_3(\xi_s+1)^3+C_5(\xi_s+1)^5)+\mathcal{O}(\xi - \xi_{s})\,,\\
	v_0(\xi) &\sim -\frac{\bar{A}(\xi_s)}{\xi - \xi_{s}} + \mathcal{O}(\xi - \xi_{s})\,, \quad & u_0(\xi - 1) &\sim A\tanh(C_1(\xi_s-1)+C_3(\xi_s-1)^3+C_5(\xi_s-1)^5)+\mathcal{O}(\xi - \xi_{s})\,.
\end{align}

From the form of the late-order ansatz~\eqref{2:ansatz}, we see that the validity of the late-order term ansatz breaks down for $\eta^2 \chi^{-2} = \mathcal{O}(1)$ as $\eta \rightarrow 0$. That is, the inner region is composed of $\xi$ such that $\eta^2(\xi-\xi_{s})^{-2} = \mathcal{O}(1)$. Correspondingly we introduce the inner scaling $\xi - \xi_{s} = \eta \overline{\xi}$. From asymptotic balancing, the appropriate rescaled inner variables are
\begin{equation}\label{2:rescaling}
	u(\xi) = -\frac{\bar{A}(\xi_s)  }{ \eta\overline{\xi}} + \hat{u}(\overline{\xi})\,,\quad u(\xi+1) =\hat{u}(\overline{\xi }+ \eta^{-1})\,, \quad u(\xi-1) =\hat{u}(\overline{\xi}-\eta^{-1})\,, \quad v(\xi) = -\frac{\bar{A}(\xi_s) }{\eta \overline{\xi}} + \frac{\hat{v}(\overline{\xi})}{\eta}\,.
\end{equation}
Retaining the leading-order terms as $\eta \rightarrow 0$, the rescaled inner equation gives
\begin{equation}
	-\frac{2\bar{A}(\xi_s) }{\overline{\xi}^3} + \diff{^2\hat{v}(\overline{\xi})}{\overline{\xi}^2}
	= -\frac{k}{c_{\epsilon}^2} \hat{v}(\overline{\xi})\,.
\end{equation}
We express $\hat{v}$ in terms of a power series
\begin{equation}
	\hat{v}(\overline{\xi}) \sim \sum_{j=1}^{\infty}\frac{v_n}{\overline{\xi}^{2j+1}}\quad \mathrm{as} \quad \overline{\xi} \rightarrow 0\,,
\end{equation}
and note that we include the leading-order singularity as part of the rescaling process \eqref{2:rescaling}. This yields
\begin{equation}
	-\frac{2\bar{A}(\xi_s) }{\overline{\xi}^3} +\sum_{j=1}^{\infty}\frac{(2j+1)(2j+2)v_j}{\overline{\xi}^{2j+3}} = -\frac{k}{c_{\epsilon}^2}  \sum_{j=1}^{\infty}\frac{v_j}{\overline{\xi}^{2j+1}}\,.
\end{equation}
By matching orders of $\overline{\xi}$, we obtain the recurrence relation
\begin{equation}\label{recur}
	v_1 = 2\bar{A}(\xi_s) c_{\epsilon}^2/k \,, \quad (2j+2)(2j+1)c_{\epsilon}^2 v_{j} = - k v_{j+1}\,.
\end{equation}
Solving the recurrence relation \eqref{recur} gives
\begin{equation}\label{series_app}
	v_j = \bar{A}(\xi_s)
     \left(-1\right)^{j+1}\left(\frac{ c_{\epsilon}^{2}}{k}\right)^{j}\Gamma(2j+1)\,.
\end{equation}
By comparing the series expression \eqref{series_app} with the inner limit of the late-order ansatz, we obtain
\begin{equation}\label{A:Lambda}
	\Lambda_s = \lim_{j\rightarrow \infty} \frac{v_j (\i \sqrt{k}/c_{\epsilon})^{2j+1}}{\Gamma(2j+1)}
	        = -\frac{\i\bar{A}(\xi_s)  \sqrt{k}}{c_{\epsilon}}\,.
\end{equation}

\section{Detailed Stokes switching analysis}\label{a:switching}

Inserting \eqref{3:series} into the governing equations \eqref{1:gov1}--\eqref{1:gov2}, we obtain
\begin{align}
	c_A^2 S''(\xi) &\sim - k R_N(\xi)\,,\label{3:RSeq1}\\
	\eta^2 c_A^2 R''(\xi) + \eta^{2N} c_A^2 v_{N-1}''(\xi) & \sim -k R_N(\xi) \quad \mathrm{as} \quad \eta \rightarrow 0\,,
	\label{3:RSeq2}
\end{align}
%as $\eta \rightarrow 0$,
where omitted terms are smaller than the retained terms in the limit $\eta \rightarrow 0$.
Noting that~\eqref{3:RSeq2} decouples from \eqref{3:RSeq1},
we study it independently. By applying the late-order ansatz~\eqref{2:ansatz} and rearranging, we obtain
\begin{equation}\label{3:Seq}
	\eta^2 c_A^2 R_N'' + k R_N \sim -\frac{\Lambda\eta^{2N} (\chi')^2 \Gamma(2N + 1)}{\chi^{2N + 1}}\quad \mathrm{as} \quad \eta \rightarrow 0\,.
\end{equation}
As we will see later, the right-hand side of \eqref{3:Seq} is exponentially small except in a neighborhood around the Stokes curve. Away from the Stokes curve, we apply the Liouville-Green method to obtain
\begin{align}
	R_N\sim C\e^{-\chi/\eta}\quad \mathrm{as} \quad \eta \rightarrow 0\,,
\label{3:WKB1}
\end{align}
where $C$ is a constant to be determined.

In the neighborhood of a Stokes curve, $R_N$ takes the form
\begin{equation}\label{3:WKB}
	R_N(\xi) \sim \mathcal{S}(\xi)\mathrm{e}^{-\chi/\eta}\quad \mathrm{as} \quad \eta \rightarrow 0\,,
\end{equation}
where $\mathcal{S}(\xi)$ is a Stokes-switching parameter and it is constant except in the neighborhood of the Stokes curve.
 Inserting \eqref{3:WKB} into \eqref{3:Seq} and rearranging, we have
 \begin{equation}
	\diff{\mathcal{S}}{\xi} \sim \frac{ \Lambda\chi' \eta^{2N-1} \Gamma(2N+1)}{2\chi^{2N + 1}}\,\mathrm{e}^{\chi/\eta}\quad \mathrm{as} \quad \eta \rightarrow 0\,.
\end{equation}
Writing $N_{\mathrm{opt}}$ in terms of $\chi$ using~\eqref{e:Nopt}, expanding the gamma function using Stirling's formula, and transforming to make $\chi$ the independent variable, we get
\begin{equation}\label{this1}
	\diff{\mathcal{S}}{\chi} \sim \frac{\Lambda\sqrt{\pi} \eta^{|\chi|/\eta + 2\omega-1}(|\chi|/\eta)^{|\chi|/\eta + 2\omega-1/2} \e^{-|\chi|/\eta}}{\sqrt{2} \chi^{|\chi|/\eta+2\omega+1}}\,\mathrm{e}^{\chi/\eta}\quad \mathrm{as} \quad \eta \rightarrow 0\,.
\end{equation}

Writing $\chi = \rho \mathrm{e}^{\i \theta}$, we transform \eqref{this1} into polar coordinates. We obtain the variations of $\mathcal{S}$ in the angular direction as
\begin{equation}
	\diff{\mathcal{S}}{\theta} \sim{\i\Lambda}\sqrt{\frac{\pi \rho}{2\eta^3}}\exp\left(\frac{\rho}{\eta}(\mathrm{e}^{\i\theta} - 1) - \frac{\i\theta \rho}{\eta} - 2 \i \omega \theta \right)\quad \mathrm{as} \quad \eta \rightarrow 0\,.
	\label{1:A}
\end{equation}
The right-hand side of~\eqref{1:A} is exponentially small in $\eta$, except in the neighborhood of $\theta=0$. Defining an inner region $\theta=\eta^{1/2}\bar{\theta}$, we find that
\begin{align}\label{this2}
	\frac{\d \mathcal{S}}{\d\bar{\theta}}\sim\frac{\i\Lambda}{\eta }\sqrt{\frac{\pi \rho}{2}}\e^{-\rho\bar{\theta}^2/2}\,.
\end{align}
By integrating \eqref{this2}, we obtain that as the Stokes curve is crossed the behaviour of $\mathcal{S}$  is
\begin{align}\label{this3}
	\mathcal{S}\sim\frac{\i\Lambda}{\eta}\sqrt{\frac{\pi }{2}}\int_{-\infty}^{\sqrt{\rho}\bar{\theta}}\e^{-s^2/2}\d s \quad \mathrm{as}\quad \eta\to0\,.
\end{align}
Evaluating the integral in \eqref{this3}, we find that the difference between the values of $\mathcal{S}$ on the two sides of the Stokes curve is
\begin{equation}
	[\mathcal{S}]_-^+ \sim \frac{\i \pi \Lambda}{ \eta }\quad \mathrm{as} \quad \eta \rightarrow 0\,,
\end{equation}
where $[\mathcal{S}]_-^+$ denotes the change in $\mathcal{S}$ as the Stokes curve is crossed from $\theta < 0$ to $\theta > 0$.
Recalling~\eqref{A:Lambda} and~\eqref{3:WKB}, we find that the exponentially small contribution from $\xi=\xi_s$ is
%given by
\begin{equation}\label{this4_before}
	[R_N]_-^+ \sim \frac{\bar{A}(\xi_s) \pi\sqrt{k}}{ c_A \eta} \mathrm{e}^{-\i \sqrt{k}(\xi - \xi_s)/c_A\eta} \quad \mathrm{as} \quad \eta \rightarrow 0\,.
\end{equation}
Recall the expression of $f(\xi)$ from~\eqref{e:senmanciu}, we conclude that $\xi_s$ solves the equation $f(\xi_s)=\mathrm{i}\pi/2$ while its complex conjugate $\xi_s^*$ solves $f(\xi_s^*)=-\mathrm{i}\pi/2$.
The exponentially small contribution from $\xi=\xi_s^*$ is given by the complex conjugate of~\eqref{this4_before}.
Therefore, the total exponentially small contribution from singularity pair $\xi_s$ and $\xi_s^*$ is
\begin{equation}\label{this4}
	[R_N]_-^+ \sim \frac{ \bar{A}(\xi_s)  \pi\sqrt{k}}{ c_A \eta} \mathrm{e}^{-\i \sqrt{k}(\xi - \xi_s)/c_A\eta} + \mathrm{c. c.} \quad \mathrm{as} \quad \eta \rightarrow 0\,,
\end{equation}
where c.c. denotes the complex conjugate. We express \eqref{this4} in terms of trigonometric functions and write $\bar{A}(\xi_s)=|\bar{A}(\xi_s)|\e^{\mathrm{i}\theta_{\xi_s}}$ to obtain
\begin{equation}
	[R_N]_-^+ \sim\frac{2|\bar{A}(\xi_s)|   \pi\sqrt{k}}{\eta c_A}
\exp\left(-\frac{\sqrt{k}\mathrm{Im}(\xi_s)}{c_A\eta}\right)
\cos\left(\frac{\sqrt{k}(\xi-\mathrm{Re}(\xi_s))}{c_A\eta}-\theta_{\xi_s}\right)\quad \mathrm{as} \quad \eta \rightarrow 0\,,
\end{equation}
where $\bar{A}(\xi_s)$ defined in~\eqref{e:barA} is determined by the coefficients $C_{2n+1}$ and the amplitude $A$ of the leading-order wave.

%%%%%%%%%%%%%%%%%%%%%%%%%%%%%%%%%%%%%%%%%%%%%%%%%%%%%%%%%%%%%%%%%%%%%%%%%%%%%%%%%%%%%%%%%%%%%%%%%%%%%%%%%%%%%%%%%%%%%%%%%%%%%%%%%%%%%%%\
\bibliography{reference2}
\bibliographystyle{plain}

\end{document}